\documentclass[useamsfonts]{pasj00}

\SetRunningHead{S.Miyazaki et al.}{Suprime-Cam}

\title{Subaru Prime Focus Camera -- Suprime-Cam --}
\author{
Satoshi \textsc{Miyazaki}\altaffilmark{1,2}, 
Yutaka \textsc{Komiyama}\altaffilmark{3},
Maki \textsc{Sekiguchi}\altaffilmark{4}, 
Sadanori \textsc{Okamura}\altaffilmark{5,6}
}
\author{
Mamoru \textsc{Doi}\altaffilmark{7,6}, 
Hisanori \textsc{Furusawa}\altaffilmark{5,3}, 
Masaru \textsc{Hamabe}\altaffilmark{8},
Katsumi \textsc{Imi}\altaffilmark{3,$*$}, 
Masahiko \textsc{Kimura}\altaffilmark{4,$\dag$},
}
\author{
Fumiaki \textsc{Nakata}\altaffilmark{5,1}, 
Norio \textsc{Okada}\altaffilmark{1}, 
Masami \textsc{Ouchi}\altaffilmark{5},
Kazuhiro \textsc{Shimasaku}\altaffilmark{5,6}, 
Masafumi \textsc{Yagi}\altaffilmark{1}, and
Naoki \textsc{Yasuda}\altaffilmark{1,2}
}
\altaffiltext{1}{National Astronomical Observatory, Mitaka,
Tokyo 181-8588}
\altaffiltext{2}{The Graduate University for Advanced Studies (Sokendai)}
\altaffiltext{3}{National Astronomical Observatory of Japan, Hilo,
Hawaii 96720, U.S.A.}
\altaffiltext{4}{Institute for Cosmic Ray Research, The University of
Tokyo, Kashiwa, Chiba 277-8582}
\altaffiltext{5}{Department of Astronomy, The University of Tokyo, Bunkyo, Tokyo 113-8654}
\altaffiltext{6}{Research Center for the Early Universe, The University
of Tokyo, Tokyo 113-8654}
\altaffiltext{7}{Institute of Astronomy, The University of Tokyo, Mitaka, Tokyo 181-0015}
\altaffiltext{8}{Department of Mathematical and Physical Sciences,
Japan Women's University, Bunkyo, Tokyo 112-8681}
\altaffiltext{$*$}{Present address: Communication Network Center
(Tsu-den), Mitsubishi Electric, Amagasaki, Hyogo 661-8661}
\altaffiltext{$\dag$}{Present address: Department of Astronomy, Kyoto
University, Sakyo, Kyoto 606-8502}

\email{satoshi@naoj.org}
\KeyWords{Instrumentation: detectors, techniques: image processing, telescopes}

\begin{document}

\maketitle

\begin{abstract}
We have built an 80-mega pixels (10240$\times$8192) mosaic CCD camera, 
called Suprime-Cam, for the wide-field prime focus of the 8.2 m 
Subaru telescope. Suprime-Cam covers a field of view 
34'$\times$27', a unique facility among the 
the 8-10 m class telescopes, with a resolution of 0.''202 per 
pixel. The focal plane consists of ten high-resistivity 
2k$\times$4k CCDs developed by MIT Lincoln Laboratory, which are 
cooled by a large stirling-cycle cooler. The CCD readout electronics 
was designed to be scalable, which allows the multiple read-out 
of tens of CCDs. 
It takes 50 seconds to readout entire arrays. We designed a 
filter-exchange mechanism of the jukebox type that can hold
up to ten large filters 
($205\times 170 \times 15$ mm$^3$). The wide-field corrector is 
basically a three-lens Wynne-type, but has a new type of 
atmospheric dispersion corrector. The corrector provides a 
flat focal plane and an un-vignetted field of view of 30'
in diameter. The achieved co-planarity of the focal array 
mosaic is smaller than 30 $\mu$m peak-to-peak, which realizes 
mostly the seeing limited image 
over the entire field. The median seeing in the $I_c$-band, 
measured over one year and a half, is 0.''61.  The PSF 
anisotropy in Suprime-Cam 
images, estimated by stellar ellipticities, is about 2\% 
under this median seeing condition. At the time of commissioning, 
Suprime-Cam had the largest survey speed, which is defined as the 
field of view multiplied by the primary mirror area of the telescope, 
among those cameras built for sub-arcsecond imaging.
\end{abstract}

\section{Introduction} 

In the mid 1980's during a concept study of Subaru telescope, 
a working group on wide-field astronomy stated
that the minimum required field of view is 0$^\circ$.5 in 
diameter. This conclusion was based on the statistics of the 
apparent size of various objects. The recommended field of view 
was realized at the prime focus. The wide-field prime focus became a 
unique feature of Subaru among very large telescopes, 
such as Keck and VLT. We designed and built a CCD mosaic camera
for the prime focus:
{\bf Su}baru {\bf Prime} Focus {\bf Cam}era, Suprime-Cam.  
This paper presents the hardware and performance of Suprime-Cam.

In order to take full advantage of the good image quality 
on Mauna Kea, we set the goal of Suprime-cam as the 
realization of natural seeing limited images over the entire 
field of view. We assumed the best natural seeing to be $0.''4$ 
based on the achievements by HRCam \citep{mcclureetal89}
and a median value of 0.''7-0.''8.
One of the most critical components regarding the image
quality of Suprime-Cam is the wide field corrector for the 
prime focus. Takeshi (2000) worked out a novel design of
the corrector to realize compact size while maintaining 
image quality. At a telescope elevation 
angle of 90$^\circ$ the diameter encircling the 80 \% 
energy of a point source is $0''.15$ at the field center 
for the design wavelength (546.1 nm) and less than $0.''3$ 
for the entire wavelength region at any position in the field of view. 
The co-planarity and stability of the cold focal plane is
another key for image quality. The fast F-ratio of 1.86 of 
the prime focus requires a tight tolerance. We have developed a new
method of mounting CCDs (subsection~\ref{mosaicing}) and a mechanically
stable frame (subsection~\ref{frame}).  

\begin{figure*}
\begin{center}
\FigureFile(150mm, 150mm){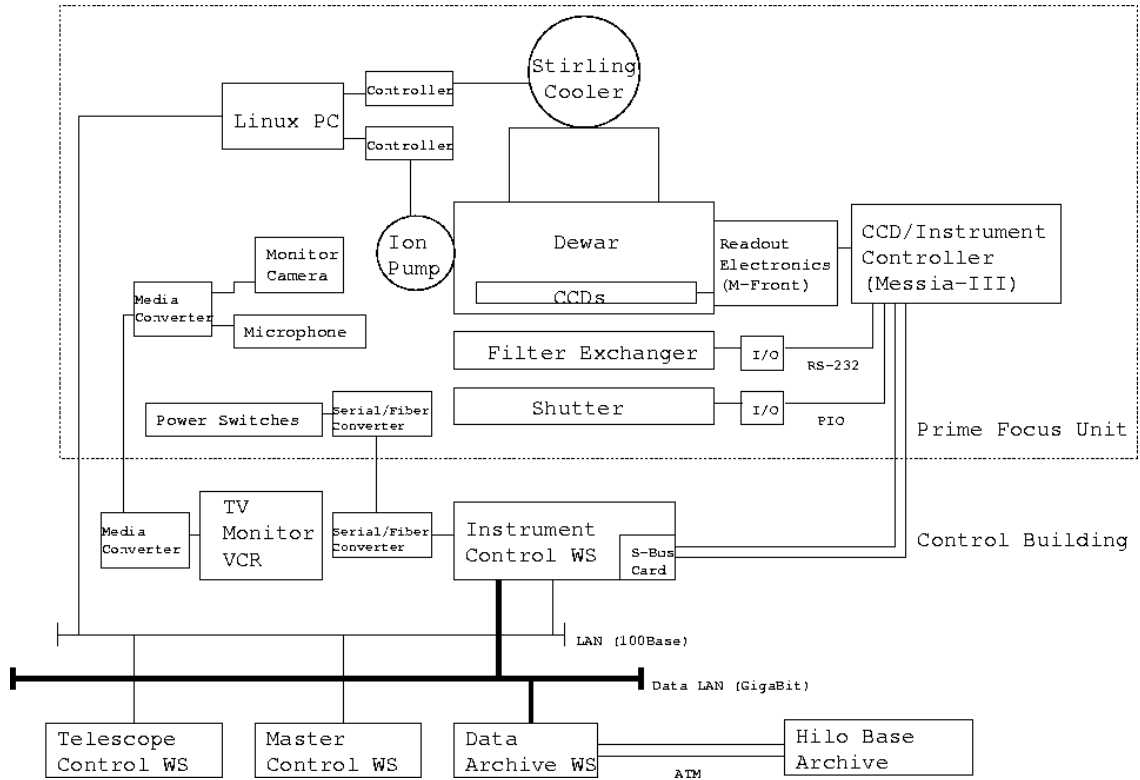}
\end{center}
\caption{Hardware block diagram of Suprime-Cam.}
\label{Fig:suphard}
\end{figure*}

The observing efficiency is also an important design goal. 
Because of the very fast F-ratio, the exposure times for broad-band 
filters are relatively short. In particular, 
they are on the order of a few to several minutes at longer 
wavelengths where the sky brightness is high. The overhead due to 
CCD readout should be minimized. CCDs that can be clocked fast 
combined with high speed readout electronics are crucial for this
purpose. All of the mechanics, including 
a shutter and a filter exchanger, must be highly reliable, since the 
camera is hard to reach once installed on the telescope. 
Accordingly, how to hold and how to exchange as many large 
(205$\times$107 mm) filters as possible is also a 
critical issue related to the observing efficiency.

WFPC2 on board the Hubble Space Telescope has indeed 
had revolutional impacts on observational astronomy. 
This is due to the superb image quality ($\leq0.''1$) 
over the modest field of view, which cannot be obtained from 
the ground in the visible wavelength, even if one employs 
adaptive optics. Computations and image simulations 
made in a design study, however, indicated that for a 
given exposure time Suprime-Cam could reach similar 
or slightly deeper limiting magnitude, though with a 
poorer spatial resolution, if all the design specifications 
were met. This was encouraging and our goal was 
the realization of stable sub-arcsec image, as good as 
$0.''4$ in extreme cases, over a more than 
100-times wider field of view than WFPC2. 
Such sub-arcsec imaging with Suprime-Cam will give many
unique opportunities to researches in various fields where 
the spatial resolution attainable only from space telescopes 
is not a critical factor.

The plan of this paper is as follows. In
section~\ref{hardwaredescription} 
we describe in detail the mechanical design of various 
components of Suprime-Cam with the intention to include 
important information useful to similar development efforts.
In section~\ref{pv} we describe the performance verification 
processes and achieved performance. Conclusions are given 
in section~\ref{conclusion}. Appendix describes the effect of 
atmospheric differential refraction.

\section{Hardware of Suprime-Cam} \label{hardwaredescription}

\subsection{Overview}
A hardware block diagram of the camera is shown in
figure~\ref{Fig:suphard}. The main science imagers, CCDs, 
are controlled by electronics called M-front and Messia-III. 
A cryogenic system, consisting of a dewar, 
a mechanical cooler and an ion pump, keep the temperature of 
the CCD at --110$^\circ$C so that the dark 
current of CCDs becomes negligible for a period on the
order of one hour. A dedicated Linux-PC controls the cooler 
and the pump by monitoring the temperature of the dewar. 
A filter exchanger and a shutter are major mechanical components of 
the camera, and are controlled through the instrument controller 
on board Messia-III. It is connected to an instrument control 
workstation (WS) located in a control building via 500 m long 
optical fiber lines. A small CCD camera and microphones are
installed inside the frame to monitor the motion of 
the mechanical parts, and those audio-visual signals are 
multiplexed and transmitted to the building via a single fiber line. 
There are several other WS's in the control room, including a master 
control WS, a dedicated WS to control the telescope and a WS for 
the data archive. The observing commands are issued from the master
control WS to others. In the following, we present descriptions 
of the hardware components of Suprime-Cam.

\subsection{CCD} \label{ccddescription}
Although the physical size of the focal plane is relatively small
($\phi$ 150 mm), thanks to the fast F-ratio, there was no CCD
available with which the entire field of view can be covered.
Therefore, mosaicing of several devices is inevitable.
At the design phase of Suprime-Cam, the largest device
commercially available was the Tektronix (later SITe) Tk2048E,
which has a 2k$\times$2k array of 24 $\mu$m pixel. The Tk2048E
used to be the most appropriate device for wide field imaging,
and has been employed by many cameras, such as the SDSS photometric
camera \citep{gunnetal98}. The Tk2048E, however, has the following
disadvantages when used for Suprime-Cam: (1) the pixel size results
in a coarse sampling of 0.''32, which is not sufficient 
for the best seeing at Mauna Kea ($\sim$ 0.''4), (2) the 
I/O bonding pads surrounding the device cause the inter-device gap 
to be larger, which is irrelevant for the relatively narrow 
physical size of the Suprime-Cam's focal plane, (3) the device 
must be read very slowly, for 3 to 4 minutes, to avoid smearing 
through losses in the charge transfer, which reduces the observing 
efficiency significantly unless the TDI mode is adopted. 

\begin{figure}
\begin{center}
\FigureFile(80mm,80mm){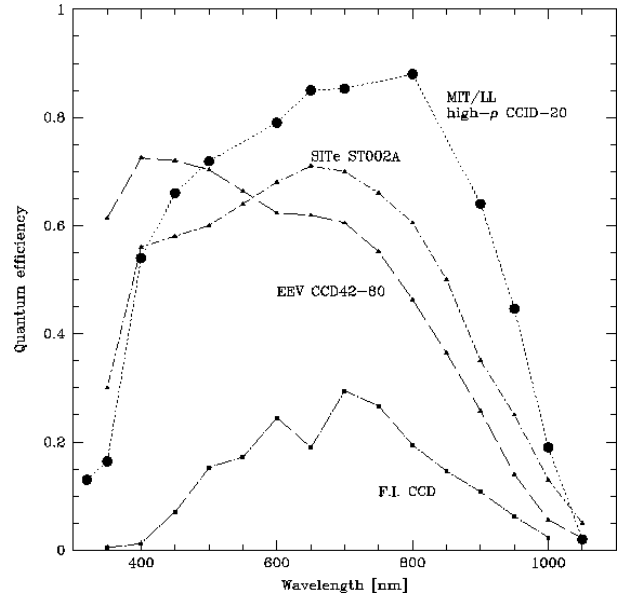}
\end{center}
\caption{Quantum efficiency of CCDs measured at NAOJ/ATC detector 
laboratory. }
\label{qecurve}
\end{figure}

In the meantime, a back-illuminated 4096$\times$2048 array of 15
$\mu$m square pixels was becoming available. We then decided to employ 
a three side buttable device of that format whose pads are 
located on a single short side so that gaps along the 
other three sides are minimized. There are three vendors 
which supplied that type of CCD:
MIT Lincoln Laboratory (MIT/LL) (Burke et al. 1998), E2V 
(the former EEV) \citep{oatesandjorden98}, and 
SITe \citep{bloukeetal98}. The SITe's ST002A has a similar quantum
efficiency (QE) as 
the Tk2048E and cannot be read out fast, either. The QE of the 
MIT/LL's CCID-20 is optimized for longer wavelength,
whereas the E2V's CCD42-80 has a high QE in shorter wavelengths 
as shown in figure~\ref{qecurve}. These two devices
can be clocked faster than ST002A and can be read out in less than 1 
minute while keeping the readout noise below $\sim$ 5 electrons. 
Because we had a high priority on the red response, we had 
adopted CCID-20 for Suprime-Cam. A sufficient number of 
CCID-20's, however, were not available before the scheduled 
first light in early 1999. We therefore mixed several ST002A's 
to the mosaic in the beginning and later made a full replacement 
of those to CCID-20's in 2000 April.

The development of CCID-20 was funded by a consortium led by 
University of Hawaii. The device was initially fabricated on an
ordinary silicon wafer, but was later fabricated on high-resistivity 
($\rho$) silicon (2-3 k$\Omega$ cm). The advantage of 
the high-$\rho$ type CCD is its thicker depletion layer, 
which improves the QE, especially at longer wavelengths 
by 15\% to 20\% compared with 
the QE of ordinary type. Another advantage of the high-$\rho$ CCD 
is the absence of interference fringing which occurs due to multiple 
reflections internal to the device. The fringing used be a real
burden of the backside-illuminated CCDs at longer wavelengths. 
The CCID-20 is the first high-$\rho$ device available for a community 
of optical astronomy. MIT/LL claims that the thickness of the 
depletion layer is about 40 $\mu$m. All of the CCDs installed in the 
camera are of the high-$\rho$ type with double layers of HfO$_2$/SiO$_2$ 
AR coating, except for one that is built on the standard-$\rho$ silicon 
with a TiO$_2$/Al$_2$O$_3$ coating. The flatness of the
device on the specification sheet given by the consortium 
was 20 $\mu$m, but the actual flatness is $\sim$ 10 $\mu$m on most of 
the devices (see figure~\ref{coplanarity-result}). Without this 
level of intrinsic flatness, we could not achieve a flat focal 
plane, as is shown in the next section.

\begin{table*}
\caption{CCD related parameters of Suprime-Cam}
\label{suprimeccdperformance}
\begin{center}
\begin{tabular}{ccccc} \hline \hline
C.F.        & Read noise        & Pixel rate  & saturation level &
dark current (--110$^{\circ}$C)  \\ \hline
2.6$\pm$0.1 e$^-$/ADU & 5-10 e$^{-}$ & 6 $\mu$s   & 92000 e$^{-}$   &
2-3 e$^{-}$/hr \\ \hline 
\end{tabular}
\end{center}
\end{table*}

Table~\ref{suprimeccdperformance} presents the CCD-related characteristics of
the Suprime-Cam measured. The Conversion Factor (C.F.) were measured at
a laboratory using MnK$_{\alpha}$ X-rays of $\;^{55}$Fe, and all the
others were measured in the telescope environment. The saturation
level in the table is the point where the response deviates by 0.5\%
from the linear response. The value is limited by the design of the 
current electronics, and is slightly lower than the value of 100000
e$^{-}$ reported by Lick observatory (Wei \& Stover 1998). 
Since the focal-plane arrays are maintained at --110$^{\circ}$C, 
the dark current is
completely negligible for the typical 5-20 min exposure.
We adopted a rather conventional pixel rate of 6 $\mu$s, 
and it takes 50 s to read out the entire array. The read noise 
is slightly higher than the best record which we obtained in the
laboratory (2-3 e$^{-}$). Although we have not fully identified 
the noise sources, we felt that the cooler and the associated power 
electronics might be the dominant causes. The sky shot noise, however,
usually dominates, and exceeds $\sim$ 50 electrons, even if one uses 
an extremely narrow band filter of $\Delta \lambda = 7$ nm with a 
reasonable length of exposure time (15 min). 

The bias level can be estimated from the over-scan-region that is 
created by clocking more times than the actual number of pixels,
and no charge is dumped into the sense node of the output amplifier. 
The variation of the bias level monitored over about one year is 
no more than 1\% around the mean thanks to the stable operating
temperature. 

A flat-field frame is usually created by stacking tens of blank 
sky images. Below 450 nm (i.e., in the $B$ band), 
the non-uniformity of the QE of the CCD
becomes apparent. The non-uniformity, called {\it brick wall
pattern}, is an imprint of the laser stepping pattern during the
backside passivation process. The degree of non-uniformity
badly depends on the device temperature. The variation of the
non-uniformity, defined as (max -- min)/mean, is about --0.5\%/deg 
at --110$^{\circ}$C (Wei \& Stover 1998). Therefore, the change is 
little if the temperature control system functions normally to 
realize $\rm \Delta T < \pm 0.3$ deg. The scattering of QE averaged 
over each device among the nine high-$\rho$ CCDs is not broad, 
3-4\% around the mean. The one standard-$\rho$ device installed 
has about 25-30\% less QE over the entire sensitive wavelength 
than others, since the AR coating of this particular device was 
not fully optimized.

The interference fringing is actually not visible on high-$\rho$ 
CCDs up to the $I_c$ band, whereas a slight sign of fringing
sometimes appears on the $z'$ band images. This is, however, 
quite small (about 0.2 \% at most) compared with sky shot noise, 
and disappears without any subtraction of a fringe frame after
dithered exposures are stacked.

\subsection{CCD Mounting} \label{mosaicing} 

\subsubsection{The requirements}
The prime-focus corrector is designed to give 
a flat focal plane where the focal depth is simply estimated 
by the pixel size (15 $\mu$m) multiplied by the F-ratio (1.86), 
$\sim \pm$30 $\mu$m (see subsection~\ref{optics}).
In order to realize the seeing-limited image, the CCD surface,
i.e., the imaging area of
the CCDs, over the entire field of view should be located within
the focal depth. We set the specification of the co-planarity at 
half of the focal depth ($=\pm$15 $\mu$m) to leave a margin for 
any mechanical errors in the flatness of the large cold plate. 
Another design requirement is the easiness of installation and 
un-installation of CCDs. This is important, since the 
best-quality high-$\rho$ devices had been hard to obtain in bulk 
in time during our development phase of the camera. 
We actually had to swap the CCDs several times for upgrades. 
Thus we have developed a new scheme for mounting CCDs to meet 
these requirements.

\subsubsection{Compensating tilt of CCDs} \label{sec_mosaic}
The CCD surface is usually tilted with respect to the bottom 
surface of the package. Thus, the height of the CCD surface from the 
bottom surface is different over a CCD chip. The height difference 
sometimes exceeds 100 $\mu$m. A conventional method to compensate for
the tilt is to insert a metal shim block between a CCD and the 
cold plate \citep{noao8k}. This requires a precise 
relative alignment of the CCD and the blocks. Furthermore, since
the heat path between the CCD and the cold plate is limited 
to the narrow channel through the metal block, which results in 
a longer cooling time, additional cold straps are necessary to make
the heat path wide. We felt that these could make the installation 
process overwhelming. We therefore developed a new method of 
CCD mounting. A detailed description of the method was first given 
in \citet{nakataetal00} during the course of the development. 
Here, we show the final results as well as a summary of 
the procedure.

\begin{figure}
\begin{center}
\FigureFile(80mm,80mm){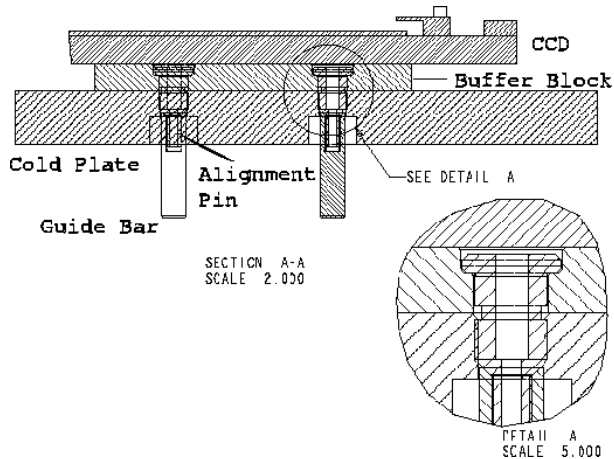}
\end{center}
\caption{
Cross sectional view of a CCD with a buffer block mounted
on the cold plate}
\label{mother}
\end{figure}

We compensated for the tilt by employing a buffer block between
a CCD and the cold plate, as shown in figure~\ref{mother}.
The  buffer block is epoxied to the CCD
with thin metal foils of the appropriate height inserted.
We chose the same material for the buffer block as
the CCD package, aluminum-nitride (AlN) ceramic, while
considering the effect of thermal expansion.
The buffer block also has alignment pins, which are
epoxied with 3M glue (2216 B/A Gray). The diameter of 
the alignment pins was originally made slightly larger than the
diameter of the holes on the cold plate. After being epoxied
to the buffer block, each alignment pin was precisely machined to
make its center coincident with the center of the hole to which it
should fit. The diameter of the alignment pin is made smaller in
this process than that of the hole by $\sim$ 30 $\mu$m so as to 
guarantee the smooth insertion.
The pins have screw-type threads that can be used to fix the buffer
block to the cold plate using nuts and belleville springs from the
opposite side.

\begin{figure}
\begin{center}
\FigureFile(80mm,80mm){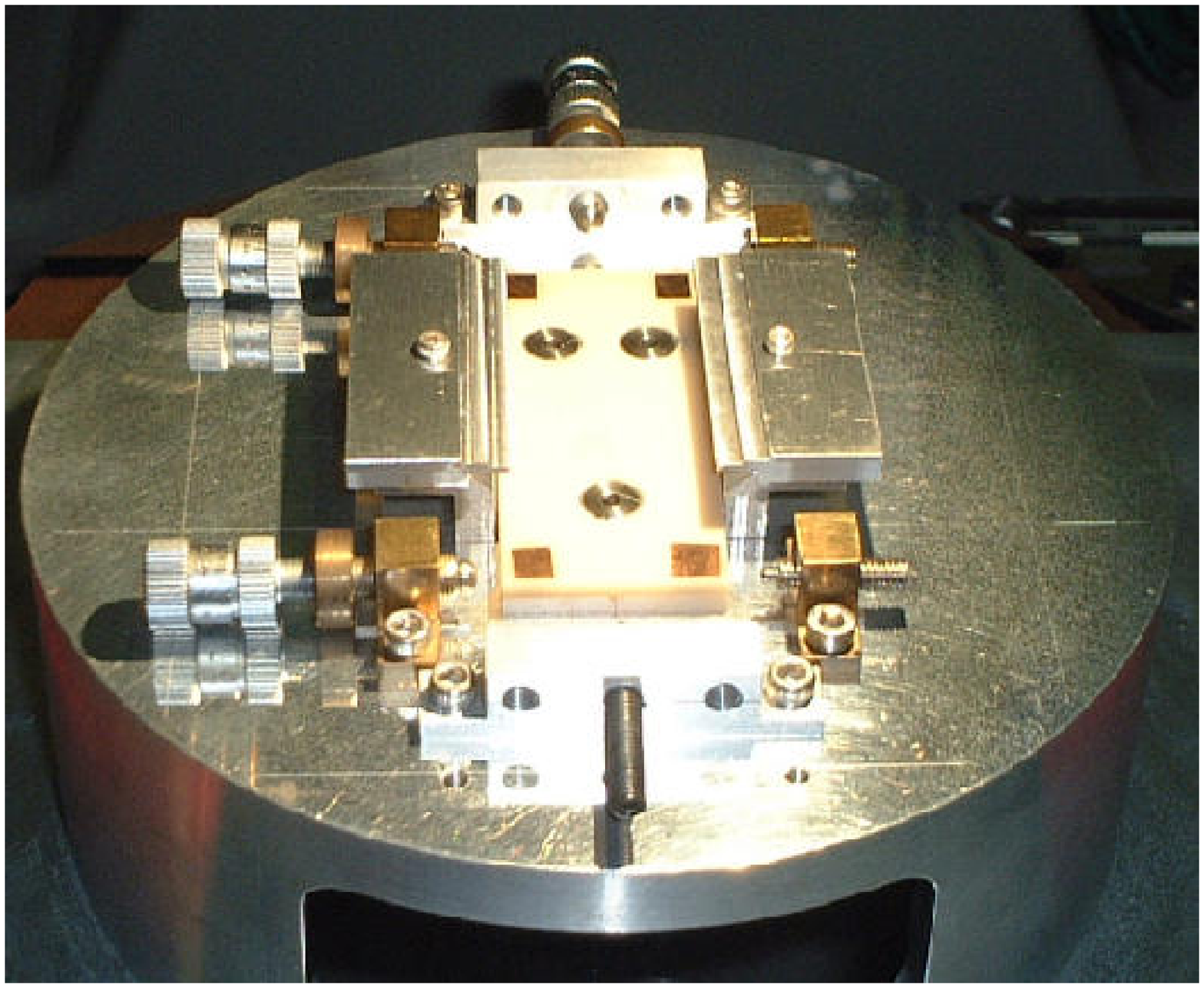}
\FigureFile(80mm,80mm){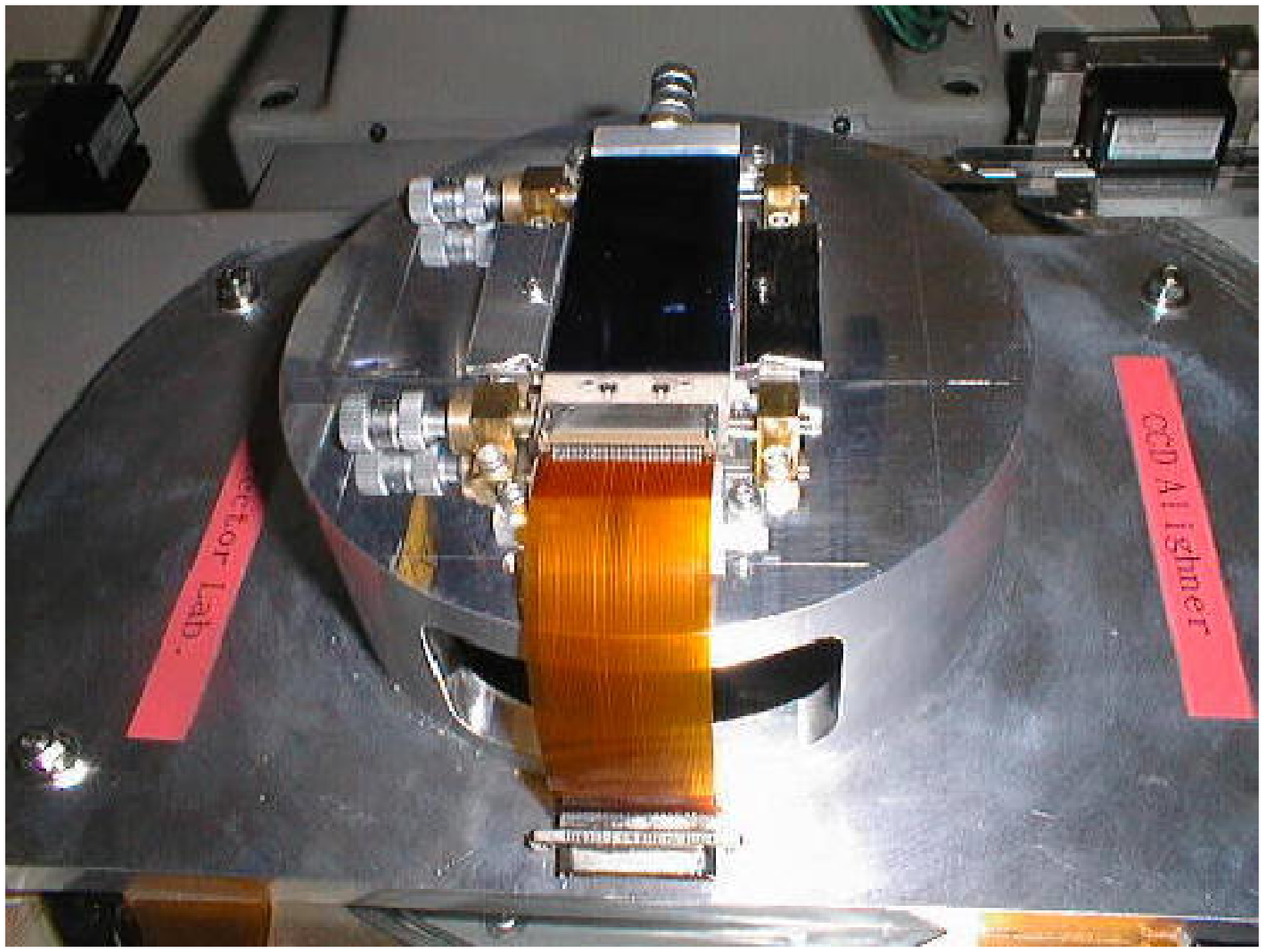}
\end{center}
\caption{Four pieces of thin metal foils are put at the four
corners of the buffer block (top). A CCD package is pushed
down by a hook on the side of the alignment bench to be glued to
the buffer block (bottom). }
\label{foil_glue}
\end{figure}

We designed and built a special working bench to
do the job, as shown in figure~\ref{foil_glue}. The bench was fixed
on an X-Y stage, which equipped with a microscope to measure the
X-Y position, and a high precision laser-displacement
meter (Keyence LC-2430) to measure the height of the imaging
area of the CCD.
First, we measured the surface profile of a CCD without the foils.
It turned out that the heights of evenly spaced 3$\times$3
points were sufficient to characterize the CCD surface,
because a height variation of smaller scale is negligible. Some 
devices have a tilt larger than 100 $\mu$m from one corner to the diagonally
opposite corner.

We then calculated the appropriate height of thin metal foils
to minimize the height variation from the fiducial plane.
The foils were inserted at the four corners of the buffer block
(figure~\ref{foil_glue}, top) and we measured the surface profile
again. If necessary, additional foils were inserted.
When the height variation became small enough (typically 10 $\mu$m
p-p), the 3M glue was filled between the CCD and the
block. The CCD was pushed down to secure the contact with the
metal foils by a hook using dips on the sides of the CCD
package (figure~\ref{foil_glue}, bottom).
When epoxying the block to the CCD, the horizontal (X-Y)
alignment was realized by pushing the CCD with screws and plungers
attached to the side of the working bench
with the position being monitored by the microscope.
After the CCD is fixed to the block, we measured
the heights to see the difference
before and after the cure. The difference was always 
as small as $\sim$ 2 $\mu$m.

Finally, we installed the CCDs with the block to the cold plate.
The cold plate was made of AlN manufactured and precisely machined 
by Mitsui Mining \& Smelting.
The X-Y location of the CCDs on the cold plate was determined by the
alignment pins attached to the buffer block and the holes on the cold
plate.
When we inserted the pins to the holes of
the cold plate, we attached cylindrical bars with a tapped hole
on top to the alignment pins, and the bars were used for a guide
to avoid butting the fragile CCDs (figure~\ref{mother}).
After insertion, the bars were removed and the nuts and springs
were used to fix the CCDs. The use of guide bars is one of the
unique features to realize the easy replacement of CCDs for a 
future upgrade.

\begin{figure}
\begin{center}
\FigureFile(80mm,80mm){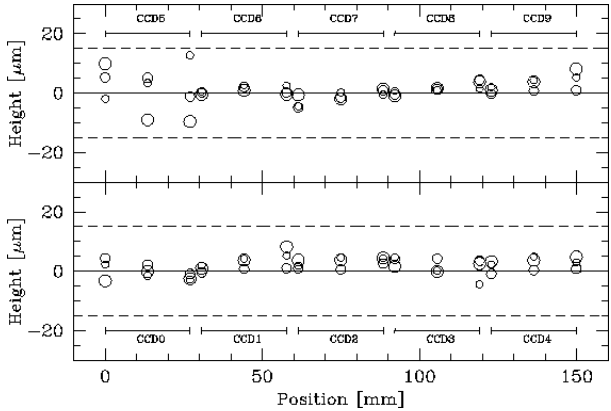}
\end{center}
\caption{Height variation of CCD surfaces after compensation.
The horizontal axis shows a distance along the short direction 
(X-direction) of the CCD and three points along the long 
direction (Y-direction) (the center and $\pm$ 26 mm) are 
measured at each X position.
The size of the symbol in the figure indicates the Y position;
the symbol becomes smaller as Y increases. The solid lines 
indicate the fiducial plane and the dashed lines correspond to
our design goal ($\pm$ 15 $\mu$m).}
\label{coplanarity-result}
\end{figure}

Figure~\ref{coplanarity-result} shows the measured
height variation of the final mosaic configuration. 
We can see that the heights of the CCD surfaces satisfy
the requirement. The distance to any point from the fiducial plane
is less than 12.7 $\mu$m. We notice that the planarity of one of
the CCD is significantly worse than those of other CCDs; 
CCD5 in the figure is apparently {\it twisted}. 
It records the largest deviation from the fiducial plane.
The typical gap between neighboring CCDs is 1.35 mm
along the short direction of CCDs and 0.81 mm along
the long direction. Each CCD has a slight rotation angle
on the order of 0.0027 radians, which is consistent with
the allowance of the fitting ($\Delta r = 30\;\mu$m) between
the alignment pins and the holes.

\subsection{CCD Readout Electronics: {\it M-Front} and {\it Messia-III}} 

\begin{figure}
\begin{center}
  \FigureFile(80mm,80mm){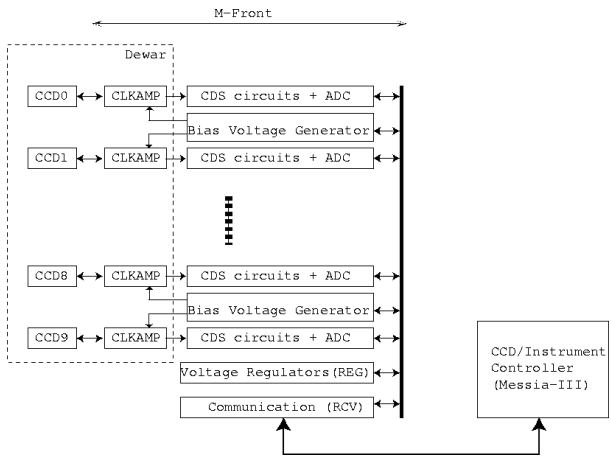}
\end{center}
  \caption{Block diagram of the CCD electronics adopted for
   Suprime-Cam; M-Front and Messia-III.}
  \label{electronicsfig}
\end{figure}

The CCD readout electronics system was originally developed in
collaboration with National Astronomical Observatory and the 
University of Tokyo. A block diagram of the electronics is
shown in figure~\ref{electronicsfig}. A DSP-based 
programmable clock generator integrated with instrument controllers is 
called {\it Messia-III} \citep{sekiguchi98}. The analog part of the CCD
electronics is called  {\it M-Front}, and its design is based on the 
electronics system of the SDSS photometric camera
\citep{gunnetal98}. Messia-III sends CCD clocks to 
M-Front and monitors the clock voltage through the I/O
port. M-Front sends 16-bit digital data back to the
frame memory on Messia-III.  We present a brief description of the 
electronics system here.

\begin{figure}
  \begin{center}
   \FigureFile(80mm,80mm){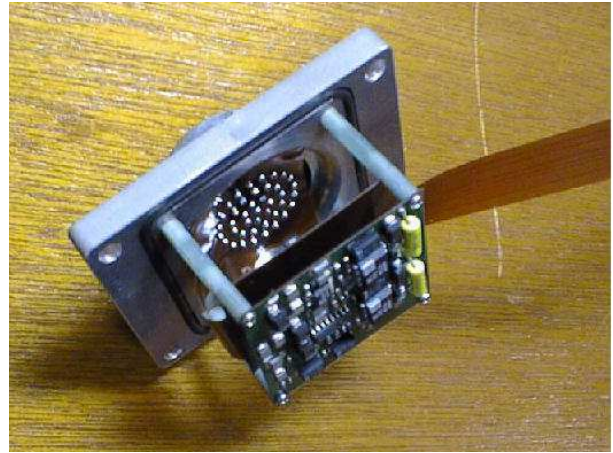}
  \end{center}
  \caption{Clkamp; a Flexible Printed Circuit board installed in the
    dewar.  The board has two pre-amplifiers (this side) and 
    analog-switch based clock line drivers (opposite side). The CCD is
    connected to the end of the flexible wiring shown to the right in
    the figure.}
  \label{flange2}
\end{figure}

Each CCD is connected to a dedicated small Flexible Printed Circuit board, 
called {\it clkamp} (figure~\ref{flange2}), which has analog-switch
based clock line drivers and two pre-amplifiers for the CCD output. 
Clkamp's are installed in the vacuum dewar to minimize the 
wiring length between the CCD and the pre-amplifier. Thanks to the 
existence of the circuits, the CCD has a low risk of electrostatic 
discharge damage once installed in the dewar. One drawback of the 
configuration is that circuit adjustments become troublesome tasks. 
Another drawback is an increase of the ungassing of these components,
which is noticeable from the longer pumping time required. 
However, once the dewar is pumped down and the CCDs are cooled,
the vacuum can be maintained by the ion pump. We therefore think 
that the ungassing is practically harmless. 

\begin{figure}
  \begin{center}
   \FigureFile(80mm,80mm){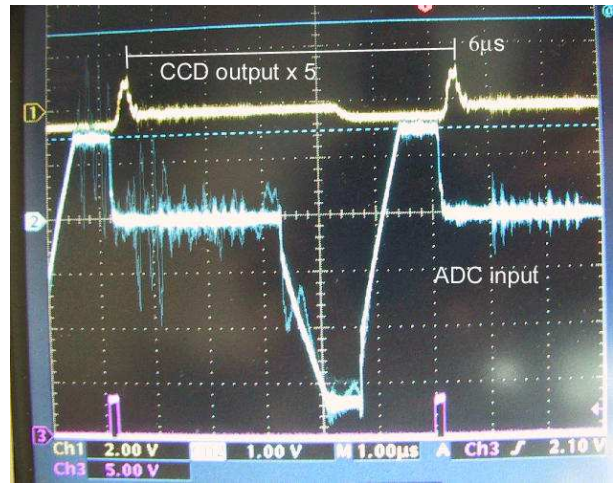}
  \end{center}
  \caption{Output waveform of the CCID20 and the output of the
    correlated double sampling of {\it M-Front} fed to ADC. It
    seems that the CCID20 can be clocked much faster because of the
    fast response of the signal. We, however, avoid critical setting 
    of the timing, since we noted that the chattering that occurred at 
    the analog switch of the CDS circuits causes an unstable output.}
  \label{waveformofccd}
\end{figure}

M-Front adopts conventional dual-slope correlated double 
sampling (CDS), whose RC constant is set at 1 $\mu$s. 
Analogic's ADC4235 (16 bit 500 kHz) is employed as an Analog to 
Digital Converter (ADC). 
The time needed to handle one pixel amounts to  6 $\mu$s, which includes 
serial clocking, the integration time of the CDS and the dead time to wait 
for a settlement of the analog switch used in the CDS. The total time to
readout the CCID20 (4k$\times$2k pixels) is thus about 50 seconds.
Figure~\ref{waveformofccd} shows the output of the clkamp 
($\times$ 10 of CCD output) and the output of the CDS fed to the ADC. 

The bias voltages for CCDs are generated by EEPROM-based DAC chips. 
M-Front has two additional boards: {\it REG}, which has power 
regulators, and {\it RCV} which receives/sends differential signals
from/to Messia-III. The size of all the boards is EuroCard 3U
(10 cm $\times$ 16 cm), except for the ADC board, which has almost 
half of the other boards. Two ADC boards are piggyback mounted 
on the bias board. All of the EuroCards are interconnected via 96 pin DIN
connectors. 

\subsection{Dewar and the Cooling System} 

\subsubsection{Mechanical design}
\begin{figure*}
\begin{center}
\FigureFile(150mm,150mm){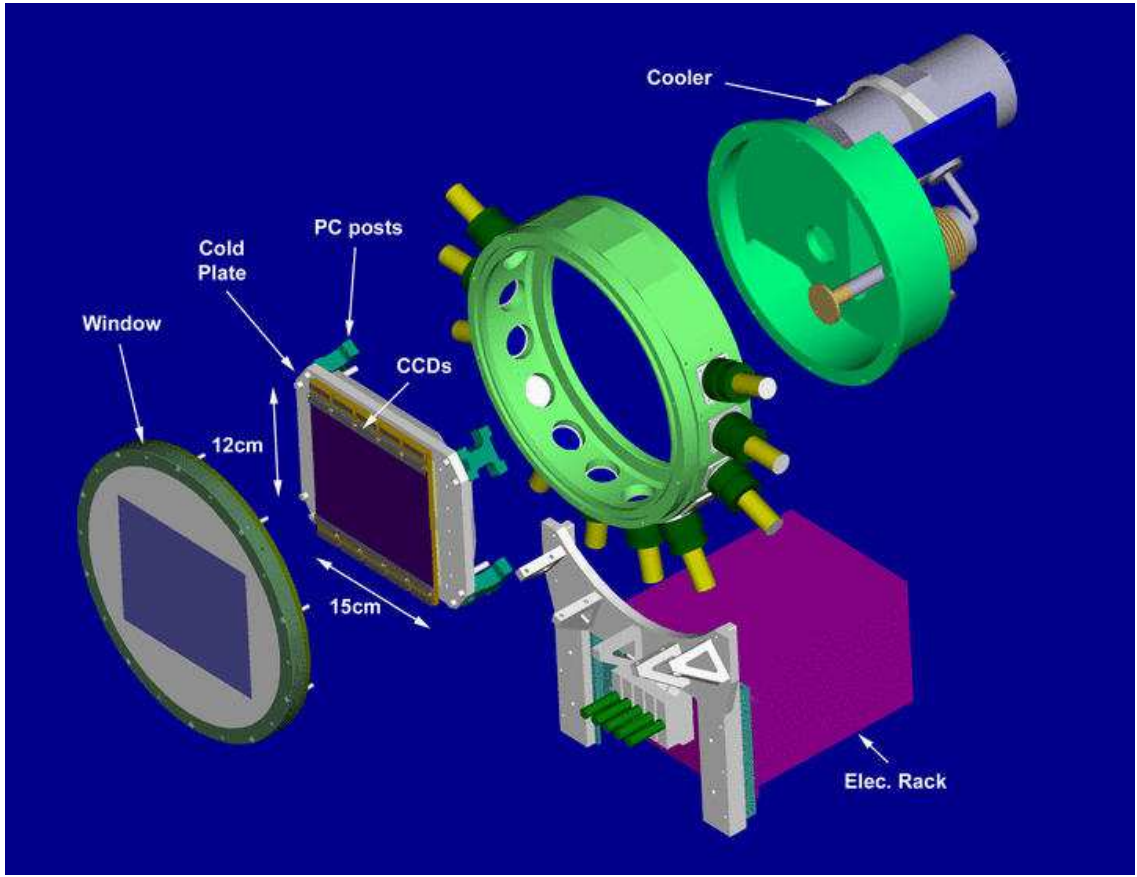}
\end{center}
\caption{Exploded view of the vacuum dewar of Suprime-Cam.}
\label{dewar_assy}
\end{figure*}

Figure~\ref{dewar_assy} shows an exploded view of the vacuum dewar of
the Suprime-Cam. CCDs are attached to an AlN cold plate 
whose size is 224 mm $\times$ 170 mm and 10 mm in thickness.
It attains 5 $\mu$m flatness on each surface and 5 $\mu$m parallelism
between two surfaces, which is the basis to make the mosaic of ten CCDs 
co-planar. Thin gold-plated radiation shields cover the plate to 
minimize the heat inflow.  

The cold plate is supported by four polycarbonate posts
(G2530; 30\% glass fiber contained) with buried taps.
Polycarbonate has a low thermal conductivity
(0.1-0.2 W m$^{-1}$ K$^{-1}$) in addition to a high Young modulus
(7-8 $\times$10$^{4}$ kg cm$^{-2}$), which makes it possible to 
keep the cold plate
isolated thermally and to hold it stably to the dewar.
These posts are designed like {\it X-shaped}, whose side and bottom
parts are trimmed so as to reduce the thermal conduction while maintaining
mechanical stiffness (see figure~\ref{dewar_assy}).

Twelve hermetic connectors are attached to the side of the
dewar for feed-through: ten for the clkamp's for the CCDs 
and two for auxiliary signals, such as temperature sensors. Cables from
these connectors are connected to the M-Front modules
attached directly to the dewar.

Anti-reflection coated 
fused silica (230 mm $\phi$ and 15 mm thickness)
is employed for the entrance window of the dewar.
The material properties of the fused silica
are listed in table~\ref{table:fusedsilica}.
The thickness of the window is determined
as follows.
For a circular disk of radius $R$ and thickness $t$, which is
supported at its circumference, the stress to the surface is 
maximum at the center and is calculated as
\begin{eqnarray}
  \sigma_{max} & = & \frac{ 3 P ( 3 + \nu )}{ 8 }
   \left( \frac{R}{t} \right)^{2},
     \label{eq:winstress}
\end{eqnarray}
where $P$ is the pressure difference between the inner and outer surfaces
of the window and $\nu$ is the Poisson ratio \citep{lifshitzandlandau}.
This stress must be less than the breaking strength, $F_{a}$ with
a safety factor of $S$ (i.e., $\sigma_{max} < S F_{a}$).
To prevent the window from cracking by tiny impact,
it is recommended that $S>3$.
Hence, we adopt $t=10$ mm for the window ($R=113.2$ mm).
In this condition, the bow at the window center, $l$, is calculated as
\begin{eqnarray}
  l &=& \frac{3 P (1-\nu)(5+\nu) }{16 E} \frac{ R^{4} }{ t^{3} },
     \label{eq:winbow}
\end{eqnarray}
where $E$ is Young modulus.
We obtain $l=$192.1 $\mu$m. We found by ray tracing
that the bow introduces little optical degradation.
Continuous flow of dry air is provided to the window surface
to prevent water condensations during observing runs.

\begin{table}
    \caption{Material properties of fused silica.}
    \label{table:fusedsilica}
  \begin{center}
    \begin{tabular}{ll} \hline\hline
      Breaking strength ($F_{a}$) & 4.892 $\times$ 10$^{7}$ \ Pa \\
      Young modulus ($E$)  & 6.966 $\times$ 10$^{10}$ \ Pa \\
      Poisson ratio ($\nu$)  & 0.17 \\ \hline
    \end{tabular}
  \end{center}
\end{table}

\subsubsection{Thermal design}
In order to maintain the dark current to be negligible,
the temperature of the CCDs
should be kept below --105$^{\circ}$C. We set the operating temperature at
--110$^{\circ}$C while allowing for some margin.
Under this condition, the dominant components of thermal inflow 
are the radiation from the window (a) and the wall of the dewar (b).
Other components include conduction through the polycarbonate
posts (c) and the wiring of electronics (d).
The convection by the remaining molecules in the dewar is less 
than $10^{-3}$ W, and can be negligible if the dewar is evacuated 
down to $10^{-6}$ Torr.
We give the thermal inflow of the four components, (a) -- (d), above.

Thermal transfer due to the radiation from a high-temperature
surface (temperature $T_{H}$ with emissivity $\epsilon_{H}$)
to a low-temperature surface ($T_{L}$ and $\epsilon_{L}$)
is calculated as
\begin{equation}
  \dot{Q}_{rad} = \sigma A (T_{H}^{4}-T_{L}^{4})
  \frac{\epsilon_{H}\epsilon_{L}}
   {\epsilon_{H}+\epsilon_{L}-\epsilon_{H}\epsilon_{L}},
     \label{eq:radiation}
\end{equation}
where $\sigma$ is the Stephan-Boltzmann constant
($5.67\times10^{8}$ W m$^{-2}$ K$^{-4}$) and $A$ is 
the area \citep{luppionoandmiller92}.
The transfer due to the conduction is 
\begin{equation}
  \dot{Q}_{cond} = \frac{\bar{\kappa} A}{L} (T_{H}-T_{L}),
     \label{eq:conduction}
\end{equation}
where $\bar{\kappa}$ is the average conductivity, $A$ the section
and $L$ the length of the material.
The total thermal inflow amounts to 7.22 W, as is given in 
table~\ref{table:thermaldesign} together with the input 
parameters of the dewar.

\begin{table*}
  \caption{Thermal inflows to the dewar.}
  \label{table:thermaldesign}
  \begin{center}
    \begin{tabular}{lll} \hline\hline
Component$^*$& Amount (W) & Input Parameters ( $T_{H}$=283 K and
$T_{L}$=163 K are assumed.)\\
\hline
(a) & 6.2 & $\epsilon_{H}$ = 1.0 (Window), $\epsilon_{L}$ = 0.5 (CCD),\\
    &       & $A$ = 224$\times$170 mm$^{2}$ \\
(b) & 0.24 & $\epsilon_{H}$ = 0.06 (Wall: Al),
      $\epsilon_{L}$ = 0.02 (Gold-plated radiation shield),\\
    &       & $A$ = 224$\times$170+2$\times$(224+170)$\times$10 mm$^{2}$ \\
(c) & 0.58 & $\bar{\kappa}$ = 0.2 W m$^{-1}$ K$^{-1}$ (Polycarbonate),\\
    &       & $A$ = 16$\times$15 mm$^{2}$, $L$ = 40 mm, 4 posts\\
(d) & 0.20 & $\bar{\kappa}$ = 420 W m$^{-1}$ K$^{-1}$ (Cu),\\
    &       & $A$ = 75$\times$18 $\mu$m$^{2}$, $L$ = 100 mm, 300 lines
    for 10 CCDs \\
\hline
      Total  & 7.22 & \\ \hline
    \multicolumn{3}{@{}l@{}}{\hbox to 0pt{\parbox{85mm}{\footnotesize
    Notes. 
    \par\noindent
    \footnotemark[$*$] See the text.
     }\hss}}
  \end{tabular}
  \end{center}
  \end{table*}

\subsubsection{Cooling system}

\begin{figure}
\begin{center}
\FigureFile(85mm,85mm){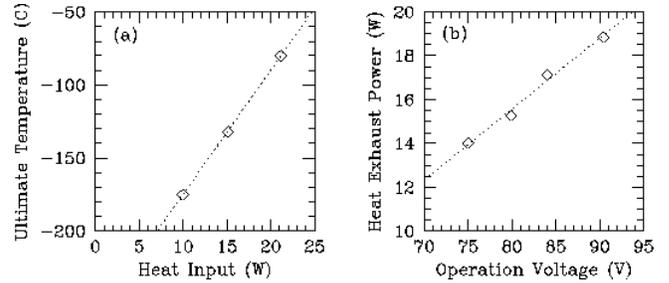}
\end{center}
\caption{(a) Ultimate temperature of the cooler head as a function 
of the heat input to the head. The cooler is operated at the maximum 
operation voltage (90 V). (b) Heat exhaust power plotted as a 
function of the operation voltage of the cooler. The cooler head 
is kept at --100$^{\circ}$C.}
\label{coolingcapacity}
\end{figure}

We adopted a stirling-cycle cooler (Daikin Industries Co., Ltd WE-5000) 
that runs stand-alone without any external equipment.
Daikin claimed that the capacity is 5 W at 80$^\circ$K; we 
tested it by changing the ohmic heat load and verified its
capacities. 
The result is shown in figure~\ref{coolingcapacity}. The WE-5000 had 
a capacity of about 16 W at --110$^{\circ}$C, where the CCDs are
operated. The value was sufficiently larger than the required 
value of 7.22 W. 

The vibration of the cooler could affect
the image quality if it induces vibration of the focal plane.
In the case that the dewar is heavier than the cooler and is 
stiffly attached to the mechanical structure of the telescope, 
it is likely that the induced vibration of the dewar is small.
Inside the dewar the cooler head is connected to a cold 
plate with a flexible copper
braid (38 mm$^{2}$ cross section and 33 mm length) so that the
transfer of the vibration becomes small. We measured the
degree of vibration of the cold plate using a laser displacement
meter (LT8110, Keyence Co., Ltd), and found that the
induced vibration is as small as 2 $\mu$m (rms) both vertically and
horizontally, which is, in fact, negligible compared with the 
pixel size (15 $\mu$m = 0.''202).

In order to maintain a sufficiently high vacuum,
charcoal is often used as a {\it getter} of molecules, such as
water.  This type of getter is, however, practically useless
unless it is cooled down to $\sim$ 80 K. 
Since we use a stirling cooler there is no place
where such a low temperature is achieved. 
We therefore adopted an ion pump 
(Noble Pump 912-7120; 20 l/s pumping speed; Anelva Co.) as 
an active getter. The pump has a finite lifetime, which is
long enough (20000-35000 hr) if the pump is operated under a
high vacuum ( $<$10$^{-5}$ torr ).

The cold plate is cooled down to --110$^{\circ}$C in 5 hours from 
room temperature. Once the system is cooled down, the power of the
cooler is regulated by monitoring the temperature of the cold plate.
Once stabilized, the cooler consumes less than 80\% of the
maximum operation power i.e., it exhausts about 11 W, which 
nearly equals the thermal inflow to the dewar. 
Under this stable operational condition, inside of the dewar
is kept below 5$\times$10$^{-7}$ torr at which point 
the thermal inflow by convection is completely negligible. 
One drawback of the stirling cooler is that the overhaul 
maintenance cycle is relatively
short, 5000 hours. It is one and a half years if we use 
the cooler for 10 days per month. The overhaul costs 
approximately \$15K, which is about one third of the price 
of the cooler.

\subsection{Filters} 
We have adopted two photometric band systems for 
broad-band filters, 
the Johnson-Morgan-Cousins system 
(Johnson \& Morgan 1953; Cousins 1978; Bessell 1990)
and the Sloan Digital Sky Survey (SDSS) system 
(Fukugita et al. 1996).
The broadband filters available at the time of this writing 
were $B,V,R_C,I_C$ of the Johnson-Morgan-Cousins system 
and $g',r',i',z'$ of the SDSS system.
We have built an SDSS $u'$ filter as well, 
but this filter has not been made available for open use
since the transmission of the prime focus corrector is quite 
low at the wavelengths of this filter. 
The transmission of the corrector is zero at $<3400$ \AA
and increases to 50\% at 3700 \AA and 80\% at 4000 \AA.
Note that the QE of the MIT/LL CCDs is also low at 
UV wavelengths: $\simeq 20\%$ at 3500 \AA and $\simeq 50\%$ 
at 4000 \AA.

The specifications for the dimensions of a filter 
are $205.0 \pm 0.5$ mm (height), 
$170.0 \pm 0.5$ mm (width), 
and $15.0 \pm 0.2$ mm (thickness), 
with a corner cut of C5-C8 (mm) for each of the four corners.
The typical weight of a filter is 1.3 kg.
A detailed description of the specifications can be found 
on-line\footnote{http://subarutelescope.org/Observing/Instruments/SCam/user\_filters.html}. 

The filters are made of one or two Schott color-glass 
elements together with two or three neutral-glass elements.
A shortpass multilayer interference film is coated 
on one glass-air surface for $B,V,R_C,I_C,g',r',i'$, 
and a bandpass multilayer interference film is 
put between two glass elements for $z'$.
Except for $z'$, the shape of the transmission 
at the short-wavelength side is determined 
by the long-pass transmission of color glass.
The physical components of the filters are summarized 
in table~\ref{filtersummary}.

\begin{table*}
\caption{Physical components of the broad-band filters.}
\label{filtersummary}
\begin{center}
\begin{tabular}{cll}
\hline
Filter & Glass (thickness in mm) & Coating \\
\hline
$B$    & GG385(3) $+$ BG40(3) $+$ BK7$^*$(3) $+$ BK7(6) &
         short-pass coating \\
$V$    & GG495(5) $+$ BK7(5) $+$ BK7(5) &
         short-pass coating \\
$R_C$  & OG590(5) $+$ BK7(5) $+$ BK7(5) &
         short-pass coating \\
$I_C$  & RG9(5) $+$ BK7(5) $+$ BK7(5) &
         short-pass coating \\
$g'$   & GG400(3) $+$ BG40(3) $+$ BK7(3) $+$ BK7(6) &
         short-pass coating \\
$r'$   & OG550(4) $+$ BK7(3) $+$ BK7(3) $+$ BK7(5) &
         short-pass coating \\
$i'$   & RG695(4) $+$ BK7(5) $+$ BK7(6) &
         short-pass coating \\
$z'$   & RG695(4) $+$ BK7(5) $+$ BK7(6) &
         band-pass coating \\
\hline
    \multicolumn{3}{@{}l@{}}{\hbox to 0pt{\parbox{85mm}{\footnotesize
    Notes. 
    \par\noindent
    \footnotemark[$*$] BK7 is neutral glass.
     }\hss}}
\end{tabular}
\end{center}
\end{table*}

Laboratory experiments have revealed that the transmission 
function at the long-wavelength side, 
which is defined by the coating, 
of the $B,V,R_C,I_C,g',r',i'$ filters,
changes by a nontrivial amount with the temperature and humidity.
That is, the cutoff wavelength 
(wavelength where the transmission is 50\% of the maximum)
becomes shorter with decreasing temperature and humidity.
This is caused by the absorption of water vapor in the air;
the maximum amount of the absorption depends solely on the temperature.
Since the absorption {\it saturates} when 
the humidity exceeds an extremely low threshold, 
the transmission of filters is expected 
to shift by the maximum amount for the temperature during the 
operation of Suprime-Cam ($\approx 0^\circ$C with a typical 
variation of less than 5 deg).
It has been found from experiments that the maximum shift of the 
cutoff wavelength from $T=20^\circ$C to 0$^\circ$C is 
$\Delta \lambda = -40$ to $-80$ \AA, depending on the filters.
In addition, the cutoff wavelength at the short-wavelength side 
(except for $z'$),
which is determined by color glass, is also found to depend 
on the temperature, though the amount of the shift is as small 
as $\sim -10$ \AA{\hspace{5pt} for $\Delta T=-20$ deg. 
Note, however, that the variation in the temperature at the filter 
during the operation of Suprime-Cam 
is typically smaller than $\sim 5$ deg.
Thus, the transmission curve of the filters does not change 
significantly during the observation.
The band-pass film of the $z'$ filter is 
put between two glass elements, not coated on an air-glass surface, 
and thus no wavelength shift is seen at either side.  

The fast F-ratio, 1.86, also changes 
the transmission functions of the filters.
The change in the part defined by the coating film 
depends on the filters, and amounts to as large as 50 \AA, 
while the change is negligibly small for the part defined 
by color glass. Note that the optics of the prime focus of Subaru is
nearly telecentric. Thus, the angle of light incident on the 
filter differs at most by only $\simeq 3$ deg over the filter 
surface. The non-uniformity of the transmission due to this amount
of difference is negligible for usual observations.

\begin{figure}
\begin{center}
\FigureFile(80mm,80mm){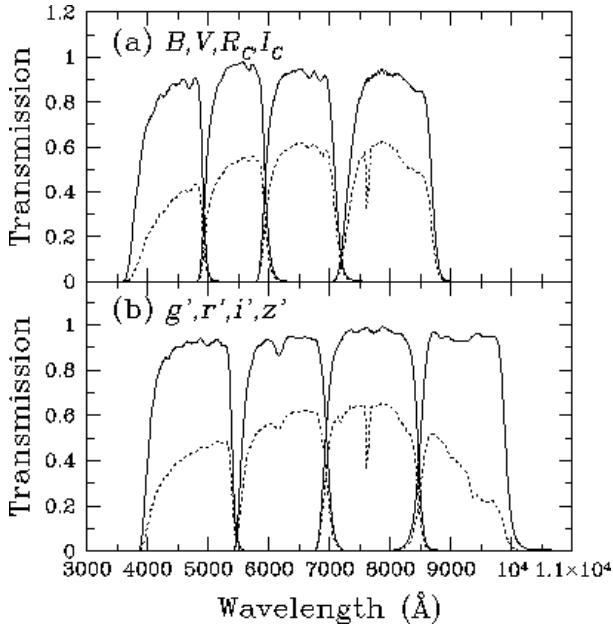}
\end{center}
\caption{Transmission curves of Suprime-Cam standard filters (solid
line). The dotted lines indicate  the combined net responses while considering
the CCD quantum efficiency, the throughput of the prime focus corrector,
the reflection of the primary mirror, and atmospheric absorption 
at $\sec z=1.2$. The deep absorption seen around 7600 $\AA$ is 
the $A$-band of molecular oxygen.}
\label{filtertransmission}
\end{figure}

For each filter, we adopt as the fiducial transmission 
the one at $T=0^\circ$C for the F/1.86 beam,
which is calculated by an extrapolation 
of the measurement at a room temperature.
The fiducial transmission functions of the eight filters 
are plotted in figure~\ref{filtertransmission} by the solid lines.
The dotted lines in the same figure indicate 
the net responses, which are defined as the combination 
of the fiducial transmissions with 
the CCD quantum efficiency,
the throughput of the prime focus corrector,
the reflection of the primary mirror, and 
atmospheric absorption at $\sec z=1.2$.
The reflection by the window glass of the dewar is
around 1\% per surface, since an anti-reflection coating
is made on both surfaces.
Table~\ref{filtercharacteristics} presents four quantities of 
these net responses:
(i) $\lambda_{\rm eff}$, 
effective wavelength against a spectrum of $f_\lambda=$const.,
(ii) $\nu_{\rm eff}$, 
effective frequency against a spectrum of $f_\nu=$const.,
(iii) FWHM, 
and (iv) $Q$ value, which is defined as 
$\int [R(\nu)/\nu] d\nu$, where $R$ is the net response.
Here, $\lambda_{\rm eff}$ is defined as 
$\int [\lambda R(\lambda)/\nu] d\lambda/
 \int [R(\lambda)/\nu] d\lambda$; 
an extra $\nu$ is inserted since the CCD counts the photon number, 
not the flux, of objects 
(cf. Fukugita et al. 1995).
Similarly, $\nu_{\rm eff}$ is defined as
$\int [\nu R(\nu)/\nu] d\nu/
 \int [R(\nu)/\nu] d\nu$. 

\begin{table}
\caption{Characteristics of broad-band filters.}
\label{filtercharacteristics}
\begin{center}
\begin{tabular}{cccrc}
\hline
Filter & $\lambda_{\rm eff}$ (\AA) & $c(\nu_{\rm eff})^{-1}$ (\AA) & 
     FWHM (\AA) & $Q$ \\
\hline
$B$    &  4478 &  4417  &  807 & 0.080 \\
$V$    &  5493 &  5447  &  935 & 0.092 \\
$R_C$  &  6550 &  6498  & 1124 & 0.103 \\
$I_C$  &  7996 &  7936  & 1335 & 0.094 \\
$g'$   &  4809 &  4712  & 1163 & 0.118 \\
$r'$   &  6315 &  6236  & 1349 & 0.127 \\
$i'$   &  7709 &  7633  & 1489 & 0.119 \\
$z'$   &  9054 &  9002  &  955 & 0.056 \\
\hline
\end{tabular}
\end{center}
\end{table}

Suprime-Cam accepts custom-made filters, such as narrow band 
filters. Any filter for Suprime-Cam should be mounted on a special 
frame of a filter exchanger, which is described 
in the next subsection. This imposes rather strict specifications on
the filters.  Furthermore, after being mounted on the frame, the filter 
should be tested mechanically in the exchanger in
advance of an observation. Those who are interested in making
custom-made filters should observe the ``Suprime-Cam filter acceptance 
policy'', which can be found at the URL above .

\subsection{Filter Exchanger, Shutter and the Control System} 

\subsubsection{Filter exchanger}
One of the difficulties in building Suprime-Cam is that
the available space at the prime focus is very limited.
On the other hand, the size of the filters is significantly
large compared to the available space. This fact did not
allow us to adopt the conventional filter
exchange mechanisms, such as a wheel-type exchanger.

\begin{figure*}
\begin{center}
\FigureFile(150mm,150mm){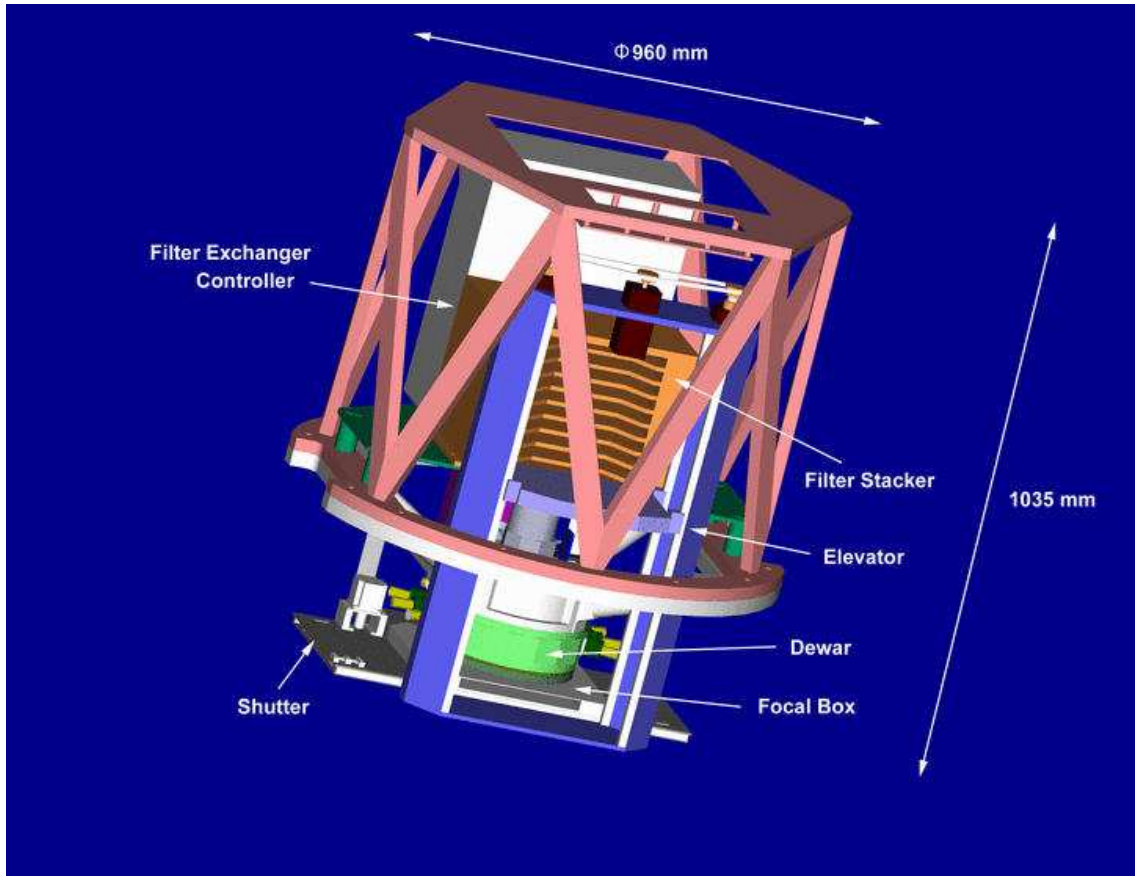}
\end{center}
\caption{Schematic view of Suprime-Cam. The major mechanical components
are shown inside.}
\label{suprime_whole}
\end{figure*}

We therefore adopted a different approach to realize the filter
exchange. The idea is based on a {\it jukebox}. 
A selected filter is first moved horizontally from the 
stored position to an elevator.
The elevator then transports the filter vertically to the focal
plane. Finally, the filter is moved horizontally again to be loaded
in the operating position. The horizontal transfer is realized 
by a rack-pinion driving mechanism and the elevator is driven 
by two ball screws and a stepping motor to achieve precise 
positioning. A filter 
holder located between the dewar and the shutter is
called the focal box. It has a positioning pin, which secures
the position of the filter within the focal box. Each filter has
a barcode label and a barcode reader installed on the elevator 
box double-checks if the right filter is loaded. The jukebox holds
up to ten filters at once.

Since each filter including the frame weighs about 3 kg,
the horizontal transfer speed of the filter should be limited 
so that the driving torque becomes sufficiently large.
It takes almost 50 seconds for a single horizontal transfer to be
completed. A series of filter exchanges (i.e., take a filter out
of the focal box, take it back to the slot of the stacker,
take the specified filter out of the slot, load it into
the focal box) takes about 5 minutes.
Note that the elevation of the telescope pointing must be high
($\ge$ 75$^\circ$ is currently required) to reduce the mechanical load 
on the horizontal transfer system during the filter exchange. 
This slewing of the telescope introduces another overhead for 
observing which is typically one minute. Therefore, it takes 6
minutes to complete the exchange. Future upgrade of the filter 
exchanger is planned to reduce these overhead significantly.

\subsubsection{Shutter}
The physical size of the aperture at the shutter amounts to 205 mm
$\times$ 170 mm.
Since iris-type shutters introduce a significant non-uniformity 
of the exposure time over such a wide field of view, we adopted a 
shutter of the sliding-door type. The space available for the camera 
in front of the focal 
plane is only 80 mm (see figure~\ref{pfunit}). The thickness 
of the shutter should thus be as thin as possible. We 
succeeded to develop a large shutter with a thickness of 15 mm.

The shutter has two sliding doors; one door covers the entire
aperture. After it slides away to open the aperture, the other door
comes in to close the aperture, which guarantees the uniformity of the
exposure time. Furthermore, no recovery time is required, since the
doors move both ways at the same speed.
Two doors are made of {\it Derlin}, which is a
solid, low-friction engineering plastic, and is driven 
individually with a timing belt and a stepper motor 
(Oriental Motor UPK543AW). The motor is controlled by pulses 
generated from the I/O port of Messia-III
(see subsection~\ref{controlsystem}) via a dedicated motor driver. 
The doors slide on rails made of the Derlin, and their 
stopping positions are determined by magnetic proximity sensors. 
It takes 1.03 s to open/close a door and the shortest
possible exposure time is set to be 1.2 s.
The exposure time is accurate to 0.02 s
and uniform to 0.3\% over the entire field of view.

\subsubsection{Control system of the mechanism} \label{controlsystem}
Messia-III has two digital signal processors (DSPs):
one for CCD control and the other
for instrument control. The stepping motors of the
shutter are controlled by the pulses from the I/O port of the DSP.
The filter exchanger is controlled by a programmable
controller (Keyence K2-500) and ladder programs. Messia-III
sends commands to/receives status from the controller through the
RS-232 communication line.

We installed a small computer system, PCI/104 based
CPU (VersaLogic EPM-CPU-6) running Linux associated
with the I/O subsystem (VersaLogic VCM-DAS-1), to manage the
cooler and the ion pump. Since the cryogenic system
are is operated continuously, even during the daytime,
we have made its control system completely independent 
so that the rest system can be shutdown if unnecessary. 

System remote reset is realized by cycling AC power to the
components. For this purpose, a remote power control
module (Black BOX SWI038A) is equipped. The RS-232C port of
the module is connected to the control workstation via a
RS-232C/optical converter (Black BOX ME570A).

Audio and visual monitoring systems are adopted in the
camera frame to diagnose the mechanism of Suprime-Cam.
Sounds of the shutter and the filter exchanger are collected by
microphones that are attached near them. They also notify observers
that an exposure starts/ends or filters are being changed.
A small video camera monitors the motion of the filter
exchanger. Since the camera is hard to reach once installed on the
telescope, these monitors are useful in the case of a possible 
system failure (e.g., a filter jam) and help to make 
quick recovery. The video composite signal and audio signals are 
converted to optical signals by a media converter 
(Force Incorporated 2792) and sent to a remote observation room.

\subsection{Frame Structure} \label{frame} 

Figure~\ref{pfunit} shows a cross-sectional view of the Subaru 
prime-focus unit (PFU) where Suprime-Cam is loaded from the top. 
The PFU is mounted and fixed on the top ring of the 
telescope. The interface between the telescope and the camera is a
torus flange located behind the prime focus by 230 mm. The camera
attached to the flange is rotated by the instrument rotator together
with an auto guider and a Shack-Haltmann sensor. On the other hand, 
the wide field corrector is fixed on the PFU. The posture of the 
PFU is controlled by a so-called 
Stewart Platform that is supported by
six hydraulic jacks. 

\begin{figure}
\begin{center}
\FigureFile(80mm,80mm){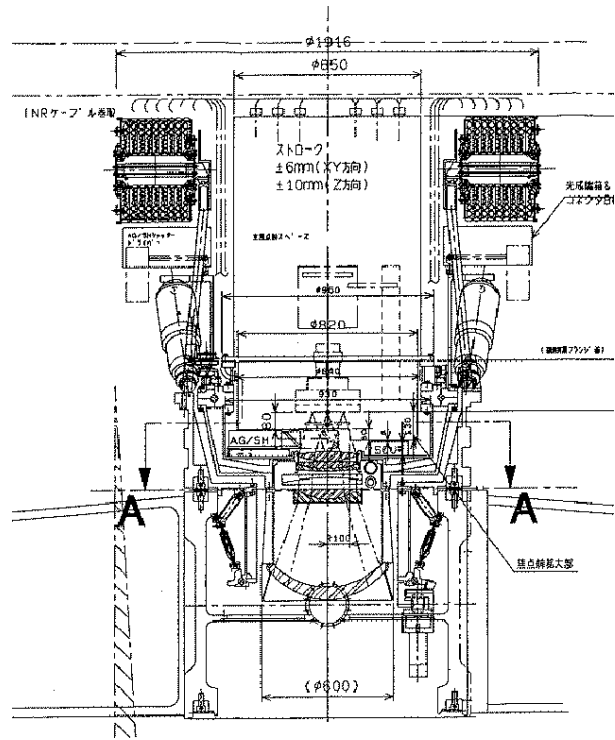}
\end{center}
\caption{Cross-sectional view of the Subaru Prime Focus Unit.}
\label{pfunit}
\end{figure}

In order to realize a seeing-limit image over the entire field of view, 
the focal-plane arrays must be placed within the thin focal
depth ($\pm$30 $\mu$m) regardless of the direction of gravity. The
tolerance of the tilt is 30 $\mu$m/100 mm $\sim$ 1', where
100 mm is the distance from the center to the edge of the arrays. 
This corresponds to a 120 $\mu$m displacement at the interface flange. 
We therefore
set the parallelism between the focal plane arrays and the interface
at 100 $\mu$m. Since reliability is our first priority, 
we did not employ any active adjustment mechanism. Instead, we 
machined each component precisely and assembled them step by step
by measuring the parallelism accurately at Ishihara Precision. 
 
Figure~\ref{suprime_whole} shows the structure of Suprime-Cam. Four
rods extending from the interface secure the vacuum dewar firmly. 
The rods are made of Invar so as to minimize the thermal deformation.
The weight of the dewar, including the cooler and electronics, 
amounts to 30 kg.  
FEM calculations showed that the tilt of the dewar is no larger than 
0.'2 regardless of the gravity direction. This was
confirmed by actual measurements in the laboratory. 

The shutter and the filter exchanger are suspended from 
the interface flange independently from the dewar.
We prepared a stage on the upper part of the flange to fix the 
filter jukebox, power units and
other auxiliary components. The upper part is covered by a truss
structure and panels, as shown in figure~\ref{suprime_whole}
(panels are not provided to show the inside).  All of the 
feed-through, including optical fibers, power and water coolant, 
are located on top of the
cover. A compact water radiator (LYTRON ES0707) is installed 
inside the camera and copes with exhaust heat of about 400 W (600 W
max) from the components.

\subsection{Optics and the Corrector} \label{optics} 
The design of the prime focus corrector is based on a three-lens
corrector for a Ritchey-Chretien hyperboloid mirror \citep{wynne65}.
\citet{nariai85} determined the positions of the three lenses
by balancing the chromatic aberration of the spherical aberration
and the image quality at the edge of the field, and
also by eliminating any higher order field curvature.

At the late stage of an engineering model study of
the Subaru Telescope, it was decided to increase the size
of the primary mirror from 7.5 m to 8.2 m.
The radius of the curvature of the primary mirror was, however, kept
unchanged so that the effects of this change on other design
specifications became minimal. Accordingly, the F-ratio of the
prime focus changed from 2.0 to 1.83, and the optical design
of the prime focus corrector had to be modified.
The trade off was between maintaining a 0.$^\circ$5 field of
view, while allowing for a slight degradation in image quality and
maintaining the image quality by reducing the field of view.
The problem was solved by a novel design in which the elements
of the atmospheric dispersion corrector (ADC) are also used in 
optimizing the corrector system 
\citep{takeshidoctorthesis,nariai94}.
Figure~\ref{Fig:corrector} shows a schematic view of the optical
components of the prime focus corrector, and note that the ADC is fully
a part of the corrector. 

The ADC consists of one plano-parallel plate and another almost
plano-parallel plate. Each of them
consists of a plano-convex lens and a plano-concave lens whose radii
of curvature are the same. The glasses of the two lenses are so chosen that
the refractive indices are approximately the same, but the dispersions are
different. These two lenses are put together with the curved
surfaces facing each other to make a plate.
When the plano-parallel plate is shifted perpendicular to the optical axis,
it behaves as an ADC. This ADC also acts as an element of the correction
of chromatic aberration of the entire system giving one degree of freedom among
the secondary spectrum, chromatic differences of aberrations, longitudinal
chromatic aberration, lateral chromatic aberration, and the power
distribution. Because of this additional one degree of freedom
coming from ADC, it was possible to maintain both the image quality and
the field of view. This new design also contributed to a reduction
of the physical size of the prime-focus corrector.

The fabrication optical data of the prime focus corrector with the 
melt-data and distances between lenses measured during assembly are given in
table 6-4 of \citet{takeshidoctorthesis}. They are given here
in table~\ref{correctoropt} for the convenience of readers.

\begin{table*}
\caption{Fabrication optical data of the primary corrector with the
melt-data and distances between lenses measured during assembly.}
\label{correctoropt}
\begin{center}
\begin{small}
\begin{tabular}{rccccllll}
\hline\hline
k & aperture & radius & distance & glass & n$_e$ & n$_{(400nm)}$ & n$_{(1000nm)}$ & Note\\
\hline
1 & 8200  & $^*$30000  & 14207.88 &         &  1  &    1    &   1     & \\
2 & 506.8 & 326.76   & 55.94   &  BSL7Y     &  1.518880 & 1.53097 & 1.50770 & \\
3 & 467   & 319.65   & 398.5995  &          &  1  &    1    &   1     & \\
4 & 275   & $^*$-4475.19983 &  15.856 & BSL7Y &  1.51854  & 1.53062 & 1.50736 & \\
5 & 251.4   & 213.36 & 65.111  &            &  1  &    1    &   1     & \\
6 & (337)  &   0     & 25.77   &  PBM5      &  1.60743  & 1.63341 & 1.58775 & \\
7 & (337)  & -897.65 & 13.962  &  BSM51Y    &  1.605811 & 1.62094 & 1.59239 & \\
8 & (337)  &   0     & 5.4995  &            &  1  &    1    &   1     & \\
9 & 258.6  & 825.93  & 14.99   &  PBM2Y     &  1.624213 & 1.65236 & 1.60319 & \\
10& 257.2  & 391.195 & 30.451  &  BSL7Y     &  1.518550 & 1.53062 & 1.50736 & \\
11& 257.2  & $^*$5355.97529  & 0.836 &      &  1  &    1    &   1     & \\
12& 257.3  & 268.24  & 48.083  & S-FPL51    &  1.498524 & 1.50774 & 1.49027 & \\
13& 254.5  & -1098.3 & 125.50955  &         &  1  &    1    &   1     & \\
14&        &         &  15.0   &  SiO$_2$   &  1.460280 & 1.47032 & 1.45056 & filter \\
15&        &         &  14.5   &            &  1  &    1    &   1     & \\
16&        &         &  15.0   &  SiO$_2$   &  1.460280 & 1.47032 & 1.45056 & window \\
17&        &         &  10.0   &            &  1  &    1    &   1     &\\
18&        &         &         &            &     &         &         & image \\ \hline
\end{tabular}

\vskip 0.1in

\begin{tabular}{rccccccc}
\hline
k & r & e$^2$ & B & C & D & D' & E \\
\hline
1 & 30000 & 1.00835  &  &  &  &  &  \\
4 & -4475.19983 & & 3.38267E-9 & -8.36303E-14 & 2.03782E-18 & -1.33838E-20 & 3.49307E-23\\
11& 5355.97529  & & 1.30769E-9 & -5.38888E-14 & -6.67037E-18 & 6.94170E-20 & -2.28747E-22 \\
\hline

    \multicolumn{8}{@{}l@{}}{\hbox to 0pt{\parbox{120mm}{\footnotesize
    Notes. 
    \par\noindent
    \footnotemark[$*$] Aspheric constants
    \par\noindent
    \footnotemark[ ] $ x =
    (h^2/r)/(1+\sqrt{1-(1-e^2)(h/r)^2})+Bh^4+Ch^6+Dh^8+D'h^9+Eh^{10}, \; h=\sqrt{y^2+z^2}. $
     }\hss}}

\end{tabular}
\end{small}
\end{center}
\end{table*}

At wavelengths of 0.40-1.0 $\mu$m for a zenith distance
of $60^\circ$ or less, the diameter which contains 
80\% of the energy of a point source at the image plane is less 
than 22 $\mu$m (0.''30) for 400 nm, 13 $\mu$m (0.''18) 
for 546.1 nm and 23 $\mu$m (0.''31) for 1000 nm over the entire
field of view. More details of including aberration diagrams, spot 
diagrams, and encircled energy diagrams based on the optical design 
are given in \citet{takeshidoctorthesis} together with verification 
measurements during the fabrication of the prime-focus corrector.

\begin{figure}
\begin{center}
\FigureFile(80mm,80mm){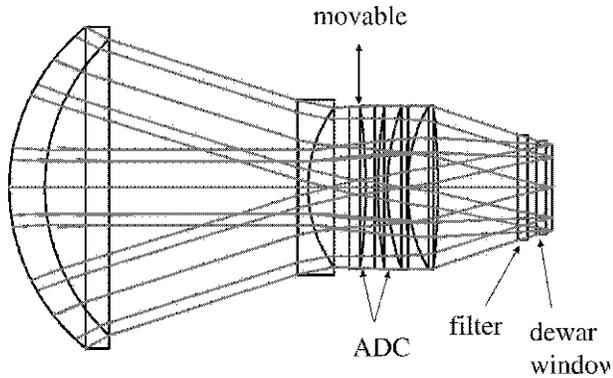}
\end{center}
\caption{Prime-focus corrector for Suprime-Cam based on a three-lens
corrector design \citep{wynne65}, but optimized with
additional optical components for ADC.}
\label{Fig:corrector}
\end{figure}

\subsection{Data Acquisition and Control Software} 

Suprime-Cam is controlled by a main program running on a SPARC
workstation with dual 168 MHz Ultra-2 CPUs, 1.4 GB memory, and a 
47 GB Ultra-SCSI RAID-5 local disk. The main program communicates with 
several other programs running in parallel: DSP programs running 
on Messia-III that control the CCD and the shutter,
ladder code programs running on the controller of 
the filter exchanger, and a status monitor (figure~\ref{Fig:suphard}).

The CCD readout is performed by a DSP program in the CCD 
controller (figure~\ref{Fig:CCDread}).
In response to a command from the main program, the DSP program 
generates a CCD clock waveform, and a stream of 
unsigned short data from the AD converter
is stored into the local memory on Messia-III.
Once the clocking comes to an end, 
a command of data transfer 
from the local memory to the shared memory of the main WS is 
initiated by the main program.
After being transferred to the shared memory,
the data are de-scrambled into 10 simple 2-dimensional 
16-bit FITS data so that each file corresponding to
a CCD chip keeps the same orientation as in the observation. 
Meanwhile, the FITS header is prepared by the main program.
The header and the data are then merged, written to the hard disk,
and sent to an archiver. While it takes 50 seconds to read out all
CCDs, there is additional overhead between exposures, including wiping
CCDs (8 s), transferring data from Messia-III to WS (52 s)
and handling the data on the WS (10 s), which is 2 
minutes in total\footnote{In 2002 August right after submission of
this paper, we upgraded the electronics Messia-III to the
new one Messia-V (Nakaya 2002 in prep.) as well as the local WS 
to an 2.2 GHz dual Pentium-based Linux box. The overhead is now 
shortened down to 58 s. Recent statistics shows that the ratio 
of total exposure time to usable observing time during the 
night reaches up to 80\% on average.}. 

The shutter is controlled by a DSP program, which uses interrupt 
signals from the Real Time Clock (RTC) on board Messia-III to 
ensure the preciseness of the exposure time regardless of the load of
main WS. The precision is less than 1 ms. 
For the filter exchanger, the main program sends command strings,
such as ``initialize'', ``set filter'', or ``restore filter'' to 
Messia-III. It simply passes the strings to the filter 
exchanger controller via RS-232C. 
When ``set filter'' command is issued, the controller also
reads the barcode of the filter which is going to be loaded 
into the focal box to check if it is the correct filter,
and returns the status to the main WS.

The status monitor of Suprime-Cam is implemented as a 
CGI (Common Gateway Interface) program.
CGI is a standard WWW technique for sending dynamic information.
The current status of the components and 
internal status of the main program are written 
into the shared memory of the main WS and constantly updated.
The CGI program reads the memory and formats the current status 
into HTML (figure~\ref{Fig:statmon}).
As the monitor program is a CGI,
any WWW browsers can be used as a remote status monitor.
We did not develop our GUI clients, because
the evolution of GUI is currently very rapid. 
Using HTTP, CGI, and HTML, which are 
stable and standard protocols for sending formatted information, 
we can easily maintain the status monitor.

During an observation, Suprime-Cam is controlled from the
Subaru Observation Software System; SOSS \citep{kosugi97,sasaki98}.
The telescope control, data archive, 
and main program of Suprime-Cam are 
commanded from a scheduler process in the master control WS in SOSS,
and act as slave processes.
Since the control is centralized in the scheduler process, 
the observation with Suprime-Cam is highly automated.
Such an automation is implemented aiming for 
future queue observations and service observations.
The scheduler also controls the synchronicity of commands
so that the dead time should be minimum.

\begin{figure}
\begin{center}
\FigureFile(80mm,80mm){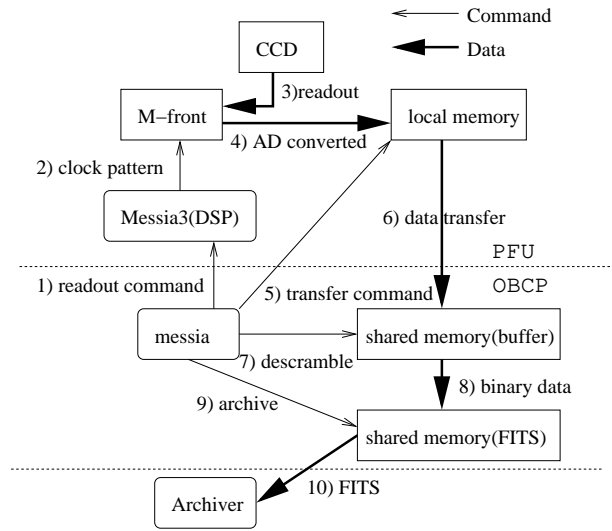}
\end{center}
\caption{Command and data-flow diagram of reading the CCD.
Command flow is expressed as the thin arrows, and data flow 
as the thick arrows.}
\label{Fig:CCDread}
\end{figure}

\begin{figure}
\begin{center}
\FigureFile(80mm,80mm){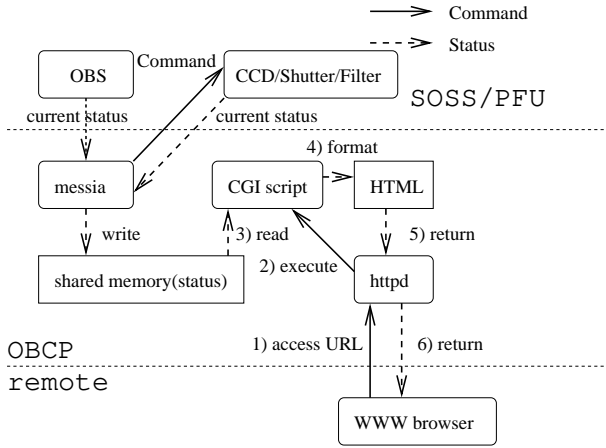}
\end{center}
\caption{Command and data flow diagram of the status monitor.
Command flow is expressed as the solid arrows, and status as the 
dashed arrows. The status from the master control WS, the CCD 
electronics, the shutter controller and
the filter-exchanger are  written into the shared memory and always 
updated by the main program. Users access to httpd running at the
instrument control WS and receive the current status in HTML format.
}
\label{Fig:statmon}
\end{figure}

\section{Performance Verification} \label{pv}
\subsection{Optical Performance} \label{opticalperformance} 
\subsubsection{Alignment of the wide field corrector} \label{alignementofcorrector}
There are three main factors that affect the image quality of 
the camera: the primary mirror, the wide-field corrector and
telescope tracking. Image degradation due to the primary
mirror is controlled to be well below 0.''2 FWHM, judging from
results of Shack-Hartman tests.
Image degradation due to the wide-field corrector is smaller
than 0.''27 FWHM, as described in subsection\ref{optics}.
The telescope tracking error has been proved to be less than 0.''07 
for a typical ten-min exposure when assisted by an auto guider
under negligible wind condition that might cause the shaking of the 
telescope. Since the quadratic sum of the three components is 0.''34,
the normal seeing (typically $\ge$ 0.''4) dominates the
image-quality degradation once all the adjustments of optics are made
appropriately.

We introduce the ellipticity of a stellar image, which is a measure
of the PSF anisotropy, which is useful to diagnose any adjustments of 
the optics, defined by
\begin{equation}
\{e_{1}, e_{2}\} = \{I_{11} - I_{22}, 2I_{12}\} / (I_{11}+I_{22}).
\label{ellipticity}
\end{equation}
The moments, $I_{ij}$, are calculated by
\begin{equation}
I_{ij} = \int d^{2}x W(\vec{x}) x_{i} x_{j} f(\vec{x}),
\end{equation}
where $f(\vec{x})$ is the surface-brightness distribution of a
stellar image and $W(\vec{x})$ is a Gaussian weighting 
function \citep{ksb95}.

The first step of the adjustment is alignment of the optical axis
of the wide-field corrector to that of the primary mirror.
If the misalignment is small, its two components, the decentering
and the tilt, can be treated separately.
The decentering causes coma aberration while the tilt causes
astigmatism.
Since the coma aberration is almost uniform over the field of view,
it is sufficient to measure it at some small portion of the field.
In fact, the decentering was measured and corrected through a 
Shack-Hartmann test performed near the field center, which controls
not only the decentering of the corrector, but also the shape of the
primary mirror.

The shear induced by the astigmatism, on the other hand, is a function
of the field position. It is efficient to use Suprime-Cam,
which covers most of the field at once, to measure the shear.
For example, if the tilt amounts to $\theta_y$ around the Y-axis, 
we note that the ellipticities of stars, $\vec{e}$ defined by
equation~(\ref{ellipticity}), can be well approximated as
\begin{equation}
\vec{e} = k\theta_y\vec{x},
\label{tilte}
\end{equation}
where $\vec{x}$ is the field position relative to the optical axis
and $k$ is a constant which is dependent on the degree of
defocus. An example of the ellipticity map is shown in
figure~\ref{ebyas}, where the mean ellipticities of stars in
11$\times$11 partitions in the field of view are shown
illustrated by ellipses.
By measuring the ellipticities of stars over the entire field we can
estimate the direction and degree of the tilt at the same time based
on equation~(\ref{tilte}). After repeating the cycle of
measurement and the adjustment several times, we were able to make
the tilt below 0.'5, which was sufficiently small.
We monitored the tilt by changing the elevation of the telescope
and no apparent change of the tilt was seen.
\begin{figure}
  \begin{center}
   \FigureFile(70m,70mm){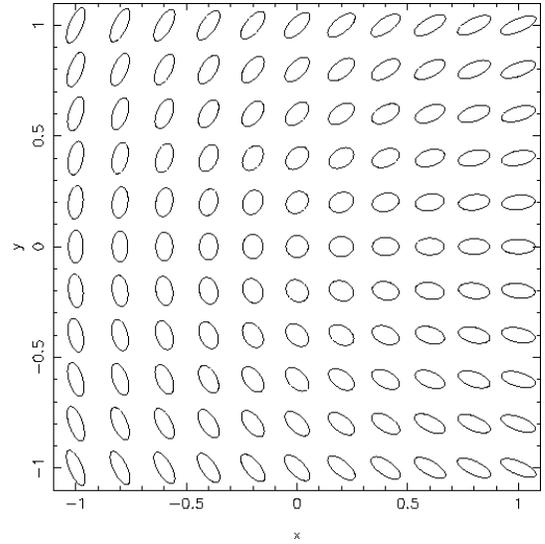}
  \end{center}
  \caption{Example of an ellipticities map ($\vec{e} = 0.5\vec{x}$; see
   the text) showing the astigmatism induced by the tilt along the
   Y-axis of the optical axis of the wide-field corrector with
   respect to that of the telescope. The mean ellipticities of
   stars in 11$\times$11 partitions in the field of view are 
   illustrated by the ellipses.}
  \label{ebyas}
\end{figure}

\subsubsection{Coplanarity of the focal plane}
We examined the coplanarity of the focal plane composed by 10 CCDs.
An exposure of 10 seconds was made through the $I_c$ band filter
under a seeing of $\sim$ 0.''4  by changing the focus position.
Figure~\ref{focus} shows the stellar image sizes averaged over each of
ten CCDs plotted as a function of the focus position.
Since the best focus position is the same for all of the CCDs,
as can be seen in figure~\ref{focus}, the co-planarity of the focal plane 
seems to be accomplished at a satisfactory level, i.e.,
any deviation from coplanarity is not visible even under superb 
seeing. The figure also shows that an image quality as good as 0.''4
can be achieved on Suprime-Cam, which proves that the three
components mentioned in subsection~\ref{alignementofcorrector} remains 
minimal, indeed. 

\begin{figure}
  \begin{center}
    \FigureFile(80mm,80mm){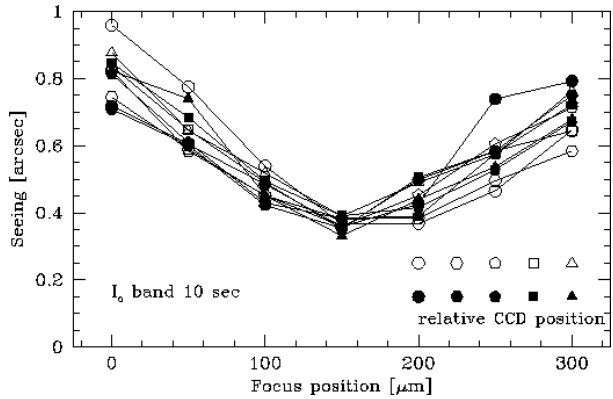}
  \end{center}
  \caption{Stellar image sizes averaged over each of the ten CCDs as a
    function of the focus position. The exposure is in the $I_{c}$ band and
    the integration time is 10 s at each position.}
  \label{focus}
\end{figure}

\subsubsection{Seeing statistics}
We monitored the seeing during the commissioning phase
over a period of one and a half years. The data are based on 
images taken for the focus adjustment procedure; the results are shown in
figure~\ref{seeingmonitoring}. The closed circles show the data taken in the
$I_c$ band, including a similar $i'$ band, which are the most frequently used
bands, and the open squares show the data of all other bands.
The median seeing is 0.''61 in the $I_c(i')$ band and 0.''69 
when all of the bands are taken into account. The best seeing that we
have ever obtained was 0.''37 in an $I_c$ band image of 180 s
exposure. The image of 0.''4-0.''45 seeing is routinely
obtained at longer wavelength (redder than $V$-band) with an 
exposure time of typically 5 to 10 min.

\begin{figure}
  \begin{center}
    \FigureFile(80mm,80mm){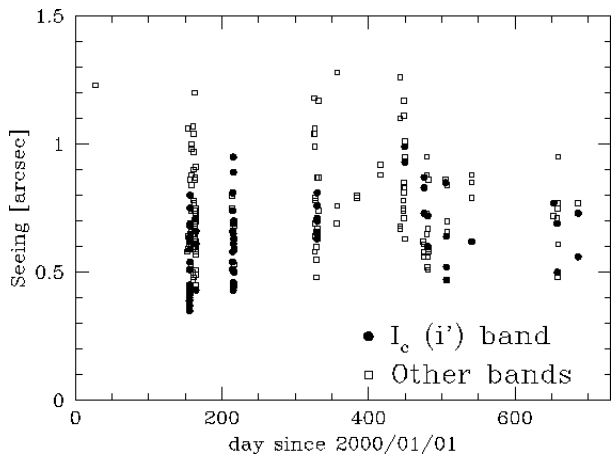}
  \end{center}
  \caption{Long-term seeing statistics of Suprime-Cam. The closed circles
    represent the data taken in the $I_{\rm c}$ or $i'$ band, which are
    the most frequently used filters, and the open squares show the data 
    for all other bands. The median seeing is 0''.61 in the
    $I_{\rm c}(i')$ band and 0.''69 when all bands are taken
    into account.}
  \label{seeingmonitoring}
\end{figure}

\subsubsection{Image quality of Suprime-Cam}
The left panel of figure~\ref{estar} shows the stellar ellipticities
measured on a certain image of 6 min exposure in the $R_c$ band.
The seeing size of the image is 0.''6, a typical value.
The mean value of the ellipticities over the field of view
is about 2\% and the standard deviation, $\sigma_e$, is 
about 1.5\%. These values are quite small compared with those of 
a camera mounted on an earlier generation telescope. For example, UH8K 
mounted on CFHT  shows 5-7\% ellipticities according to 
\citet{cfh-cosmicshear}. This demonstrates that we achieved 
sufficiently good optical and mechanical adjustments 
of the telescope/camera system.

\begin{figure}
\begin{center}
  \FigureFile(75mm,75mm){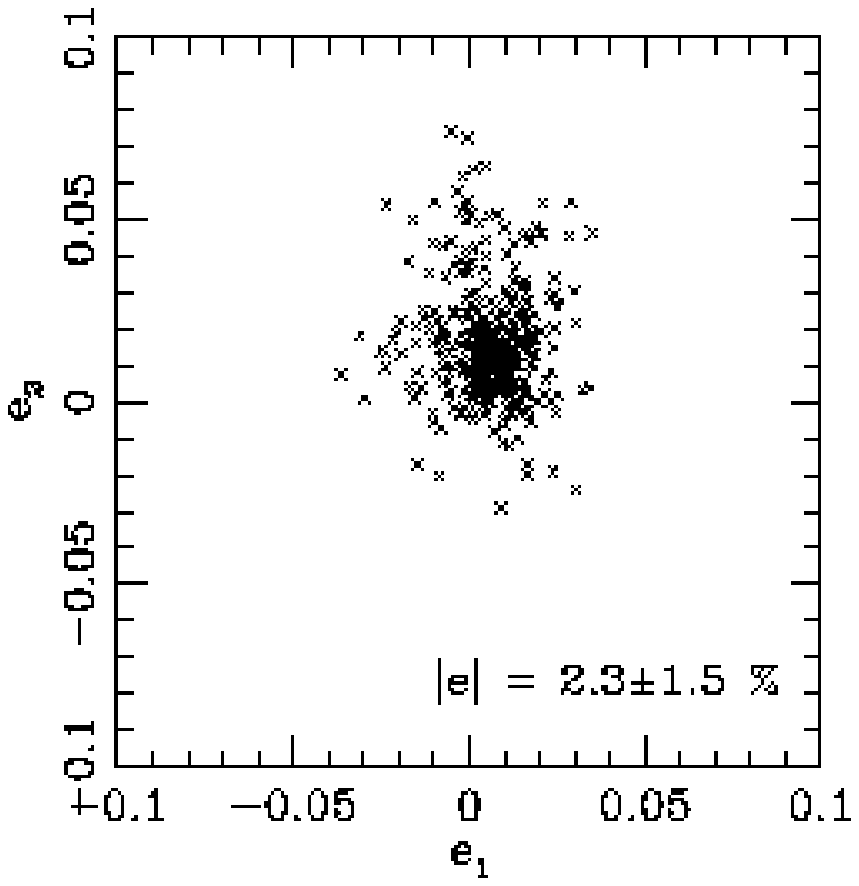}
  \FigureFile(70mm,70mm){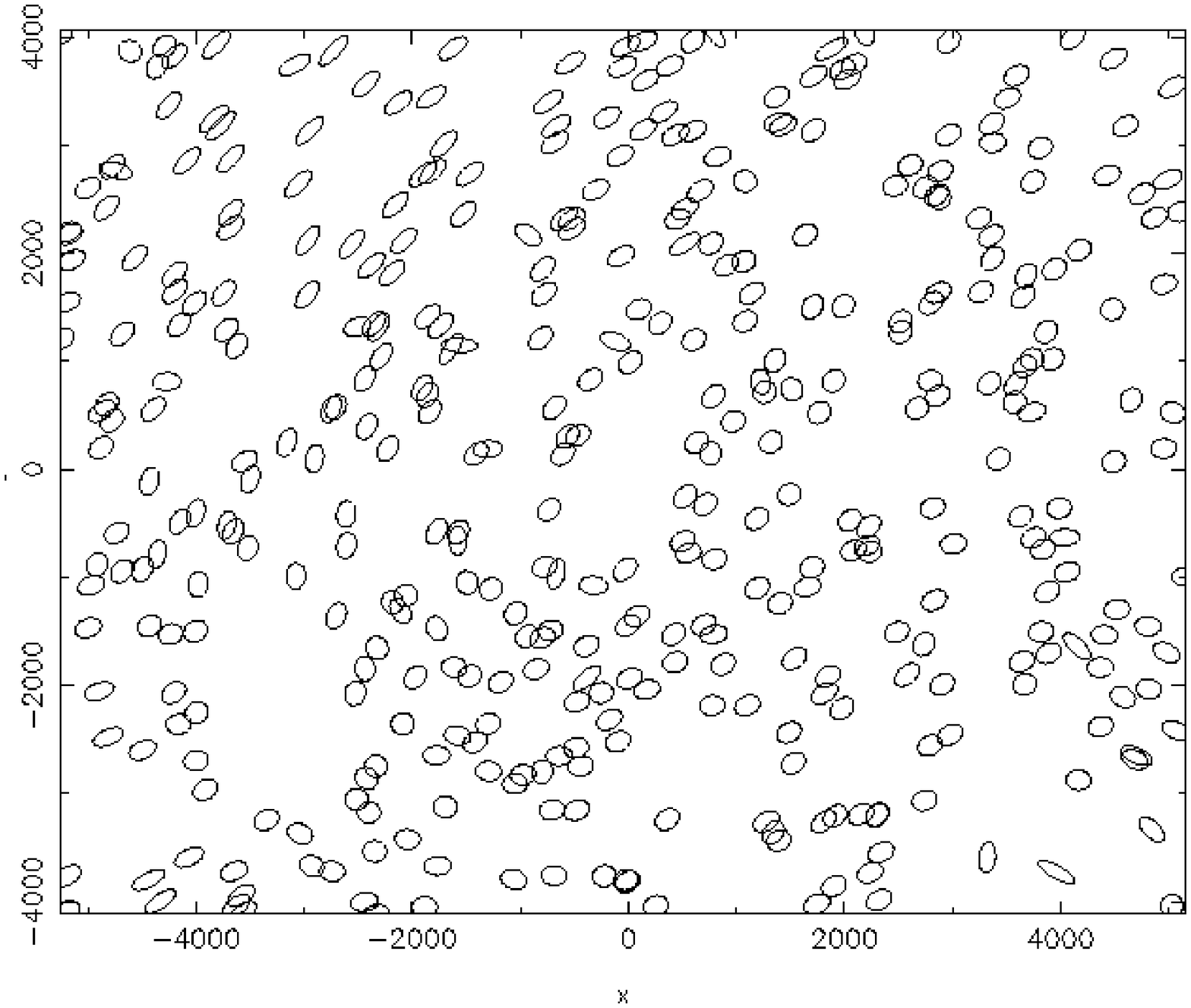}
\end{center}
  \caption{Ellipticities of stars evaluated on a certain single image 
    of 0.''6 seeing (top) and the ellipticity map
    (bottom) where the ellipticities are ten-times exaggerated for clarity.}
  \label{estar}
\end{figure}

\subsubsection{Deriving the distortion parameters}
A science exposure usually consists of several stacked images.
Image stacking is not a trivial procedure because of the large
distortion of the optics. Here, we describe how to derive the 
distortion parameters. We employ a geometrical model of field 
distortion using a 4th-order polynomial function,
\begin{equation}
\frac{R - r}{r} = ar + br^2 + cr^3 + dr^4,
\end{equation}
where $R$ and $r$ are the distances from the optical axis in units of
pixel on the CCD coordinates and the celestial coordinates, respectively.
The displacement and the rotation of each CCD from its nominal position,
$(\Delta x, \Delta y, \Delta \phi)_c$, are set as free
parameters. We change the telescope pointing typically by 1'-2' 
between successive exposures.
The offset and the rotation of the telescope pointing between
the exposures, $(\Delta X, \Delta Y, \Delta \Phi)_e$, are set
as free parameters as well. All of these parameters can be
determined by minimizing the distance of the same stars identified 
on different exposures. The residual of the distances is a measure
of the error of this mosaic-stacking procedure. The rms value of 
the residuals is typically about 0.5 pixel.

The distortion parameters determined by this procedure are
summarized in table~\ref{distparam}. Figure~\ref{distortion} shows 
the distortion computed based on our model with the
parameters given in table~\ref{distparam} compared with that 
calculated by a ray-tracing program. The measured distortion agrees 
quite well with the computation, which suggests that the solution of 
the geometry is reliable. The displacement ($t$) and the rotation ($\phi$) 
of each CCD are consistent with the allowance of fitting 
($\Delta r = 30\;\mu$m) between the alignment pins and the holes of 
the CCD mounting (subsection~\ref{sec_mosaic}), and are found 
to be stable ($dt \sim 4\;\mu$m, $d\theta \sim 0.05$ rad for 
about a year), which is natural. 

\begin{table}
\caption{Parameters of the optical distortion.}
\label{distparam}
\begin{center}
\begin{tabular}{cccc}
\hline
 $a$ & $b$ & $c$ & $d$ \\ \hline
7.16417$\times 10^{-8}$  & 3.03146$\times 10^{-10}$ & 5.69338$\times 10^{-14}$ & -6.61572$\times 10^{-18}$ \\
\hline
\end{tabular}
\end{center}
\end{table}

\begin{figure}
  \begin{center}
  \FigureFile(80mm,80mm){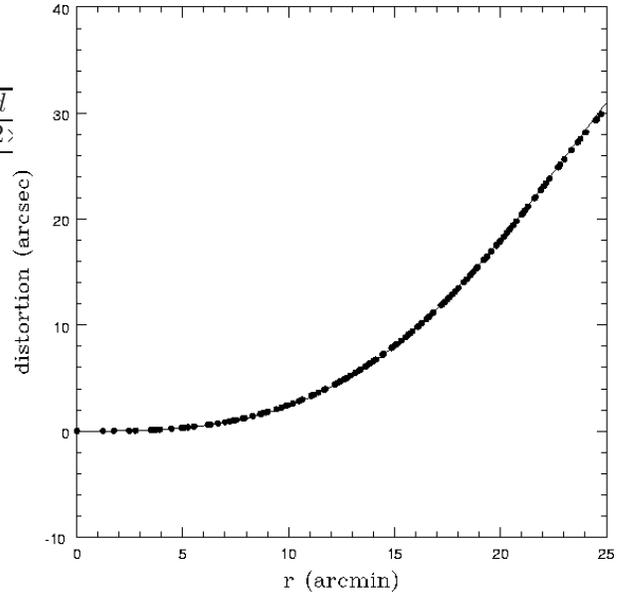}
  \end{center}
  \caption{Optical distortion of the prime-focus corrector. The solid
  circles represent the distortion calculated by ray-tracing and the 
  solid line shows the geometrical solution of Suprime-Cam (see text). Both
  agree quite well, which implies the reliability of the solution. 
  }
  \label{distortion}
\end{figure}

\subsection{Limiting Magnitude} 

\subsubsection{Magnitude zero points}

We derived the magnitude zero point, i.e., 
the $AB$ magnitude corresponding to a flux of 1 ADU per second,
for the $B,V,R_C,i'$, and $z'$ bands using data of 
photometric ($B,V,R_C$) and spectrophotometric ($i',z'$)
standard stars taken in the commissioning phase 
of Suprime-Cam.
Table~\ref{throughput} presents the results at an airmass of 
$\sec z=1.2$, together with predictions for $\sec z=1.2$
calculated with the system throughput we evaluated, which are
also given in the table.
The typical uncertainties in the observed and predicted values 
were estimated to be $\sim \pm 0.05$ and $\pm 0.1$ magnitude, 
respectively. It is found that the predictions agree with the observed 
values with discrepancies of less than $\pm 0.07$ mag. 
This demonstrates that we have correctly evaluated 
all of the factors, including the telescope optics, 
that contribute to the total throughput of Suprime-Cam.

\subsubsection{Limiting magnitudes}

We estimated the limiting magnitudes for a 1-h exposure 
for the $B,V,R_{\rm c},i'$, and $z'$ bands 
based on deep imaging data taken during the commissioning phase.
We used data of the Subaru Deep Field 
[SDF; a blank field centered at 
RA(2000)$=13^{\rm h}24^{\rm m}21.^{\rm s}4$,
DEC(2000)$=+27^\circ29'23''$] for $B,V,i',z'$, 
and those of CL$1604+43$ 
a distant cluster at $z\simeq0.9$) for $R_C$. 
The SDF is one of the two deep fields observed with Suprime-Cam 
to study the properties of distant field galaxies.
These data were taken on dark, mostly clear nights.
Their image quality was on the average good, with
a seeing size of 0.''6-0.''8.

The estimated limiting magnitudes in the $AB$ system are summarized 
in table~\ref{limitmagnitude}. 
Here, the limiting magnitude is defined as 
the brightness corresponding 
to $3\sigma_{\rm sky}$ on a $2''$-diameter aperture, 
where $\sigma_{\rm sky}$ is 
the sky noise measured as the standard deviation. 
Column 2 of table~\ref{limitmagnitude} shows the
limiting magnitude for a 1-h exposure scaled from
the raw value for the exposure time shown in the parentheses
(column 3) made at the field given in column 4. 

For the reader's reference, we plot in figure~\ref{galaxynumbercount} 
the galaxy number counts for $B,V,R_C,i',z'$ 
derived from the SDF data. 
The net exposure time of $R_C$ for the SDF was 73 min.
The solid circles indicate the counts in the SDF, 
where incompleteness of detection at faint magnitudes 
has not been corrected.
The counts at brighter than $\sim$18-20 mag are not plotted 
because images of such bright galaxies often suffer from saturation
and, thus, were removed in our analysis.
At the faintest magnitudes, 
the observed counts reach 
$\simeq 2\times 10^5$ ($B,V,R_C,i'$) 
and $1\times 10^5$ ($z'$) gals mag$^{-1}$ deg$^{-2}$.
The turnovers seen at the faintest magnitudes are artifacts 
due to the incompleteness of detection. 
The open circles in panels (a), (b), and (d) 
show the counts for $B450$ (a), $V606$ (b), and $I814$ (d) 
in the Hubble Deep Field North (HDFN, Williams et al. 1996).
The counts in the SDF agree reasonably well with those in 
the HDFN. Note that the actual number of galaxies exceeds
100000 per field of view ($R_c < 27$, 73 min exposure),
which demonstrates that Suprime-Cam substantially improves the
efficiency of statistical studies of faint galaxies.

\begin{table}
\caption{System throughput and magnitude zero points.}
\label{throughput}
\begin{center}
\begin{tabular}{lccccc} 
\hline
Band      & $B$ & $V$ & $R_C$ & $i'$ & $z'$ \\ \hline
  & \multicolumn{4}{c}{Zero point ($AB$  mag)$^*$}\\
\hline
Measured  & 27.40 & 27.54 & 27.70 & 27.92 & 27.05 \\ 
Calculated  & 27.42 & 27.58 & 27.70 & 27.85 & 27.03 \\ \hline
  & \multicolumn{4}{c}{Throughput}\\
\hline
Mirror    & 0.91 & 0.90 & 0.89 & 0.86 & 0.90 \\
Corrector & 0.85 & 0.92 & 0.93 & 0.92 & 0.85 \\
CCD       & 0.61 & 0.75 & 0.84 & 0.85 & 0.54 \\
Filter    & 0.88 & 0.97 & 0.94 & 0.96 & 0.93 \\ \hline

    \multicolumn{6}{@{}l@{}}{\hbox to 0pt{\parbox{85mm}{\footnotesize
    Notes. 
    \par\noindent
    \footnotemark[$*$] Typical airmass (sec z) of 1.2 is assumed.
     }\hss}}

\end{tabular}
\end{center}
\end{table}

\begin{table}
\caption{Limiting magnitudes of Suprime-Cam$^*$. }
\label{limitmagnitude}
\begin{center}
\begin{tabular}{cccc}
\hline
Band & $AB$ mag (1hr)  & $AB$ mag (exp.time) & Field \\
\hline
$B$  &  27.2 & 27.7  (151 min)     & SDF \\
$V$  &  27.1 & 27.4  (109 min)     & SDF \\
$R_C$&  26.8 & 26.4  (\hspace{5pt}30 min) & CL1604 \\
$i'$ &  26.5 & 26.8  (103 min)     & SDF \\
$z'$ &  25.9 & 26.0  (\hspace{5pt}68 min) & SDF \\
\hline
    \multicolumn{4}{@{}l@{}}{\hbox to 0pt{\parbox{85mm}{\footnotesize
    Notes. 
    \par\noindent
    \footnotemark[$*$] These data were taken on dark, mostly clear nights. Their image quality was on the average good, with a seeing size of 0.''6-0.''8.
     }\hss}}

\end{tabular}
\end{center}
\end{table}

\begin{figure}
\begin{center}
\FigureFile(80mm,80mm){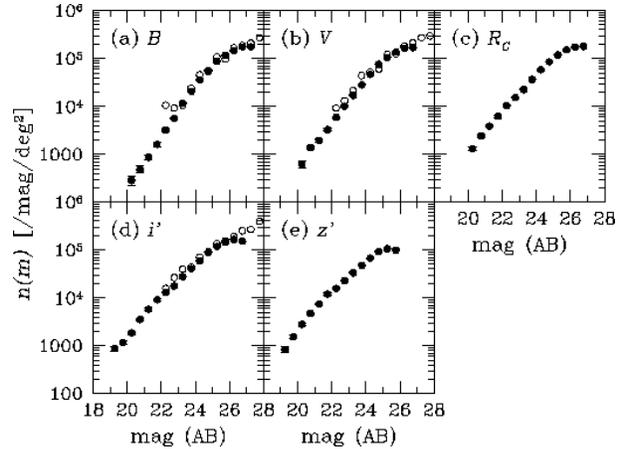}
\end{center}
\caption{
Differential galaxy number counts for $B,V,R_C,i',z'$
in the data of the Subaru Deep Field (solid circles).
Any incompleteness of detection at faint magnitudes
has not been corrected.
Counts at brighter than $\sim$ 18-20 mag are not plotted,
because such bright galaxies are often saturated,
and were, thus, removed in our analysis.
The open circles in panels (a), (b), and (d)
show the counts for $B450$ (a), $V606$ (b), and $I814$ (d)
in the Hubble Deep Field North \citep{williams96}.
}
\label{galaxynumbercount}
\end{figure}

\subsection{Comparison with Other Systems} 
The observing time of large-aperture telescopes (D $>$ 4 m) 
was once thought to be devoted primarily to spectroscopic 
observations, since imaging can be made with smaller telescopes.
Wide-field imaging with large telescopes, however,
has gradually drawn astronomers' attention since several pioneering
discoveries were reported, including weak lensing induced by
dark matter \citep{tysonetal90}, Lyman break galaxies 
\citep{steideletal96}, Lyman alpha emitters \citep{cowieandhu98} 
and high redshift type Ia supernovae \citep{perlmutteretal97}. 
The advent of large format CCDs ($\sim$4k$\times$2k pixels) around 
1994 made the efficiency of imaging observations quite high, 
and interests in imaging with large telescopes became much stronger.

Table~\ref{mosaiccomparison} presents comparisons of Suprime-Cam with
other wide-field mosaic cameras. 
In the table, $A$ shows the light-collecting power of the telescope 
measured by the area of the primary mirror, and $\Omega$ is
the field of view in square degree. Thus, $A\Omega$ represents a 
measure of the survey speed of the camera, although the exact survey 
performance depends on numerous factors, including the camera duty 
cycle, available number of nights for the survey and so on, 
which we do not deal with here.
The SDSS photometric camera is certainly the most powerful 
instrument in terms of $A\Omega$ although deep survey is not 
appropriate for the camera because of the moderate seeing 
($\sim$ 1.''2) of the site. In fact, the SDSS camera has a 
relatively coarse resolution of 0.''4 /pixel optimized 
for the wide survey to which it is dedicated. The largest aperture 
of Subaru gives Suprime-Cam the highest $A\Omega$ among the 
cameras built so far for the large telescopes that
pursue objects as faint as possible. The exposure time one needs 
to reach up to a given magnitude is shorter for larger apertures. 
Thus, Suprime-Cam is the most powerful camera for detecting faint 
transient objects. The excellent image quality demonstrated in
subsection~\ref{opticalperformance} is also a distinctive feature of 
Suprime-Cam. 

Cameras with a much wider field of view are being built
or planned, as listed in table~\ref{mosaiccomparison}.
These are motivated by a widely accepted consensus that
the progress of observational astronomy is simply limited by
the available telescope time. {\it MegaCam} \citep{megacam00},
which is being built for the 3.6 m CFHT, and is scheduled for 
the first light in 2002, has an $A\Omega$ value 
comparable to that of Suprime-Cam because of the wider 
field (1 deg$^{2}$). \citet{kaiseretal02} worked out a new scheme 
for a small 
telescope array in cooperation with innovative orthogonal 
transfer CCDs \citep{tonryetal02} to realize high-resolution 
wide-field imaging. The project is funded and four 
telescopes will be built by 2006. \citet{dmt01} are proposing 
Large-aperture Synoptic Survey Telescope (LSST), which is 
an 8.4 m telescope with about 7 deg$^{2}$ FOV dedicated for only wide 
field imaging surveys. To remain competitive in this new era of 
wide-field imaging, an upgrade of Suprime-Cam is crucial in the very 
near future. For example, a widening of the field of view up to 
4 deg$^{2}$ would make $A\Omega$ about 207, which is comparable to 
that of LSST. This is no doubt challenging and requires a 
complete re-design of the prime focus
unit and the corrector. Recent optical design shows that the new 
corrector with the first lens of 1.5 m in diameter (d) can 
realizes an image quality better than 0.''3 (diameter of 80\% 
encircled energy) over the field of view
if we allow field curvature of the focal plane and only consider
$\lambda \ge$ 550 nm. We are currently trying to reduce the size of
the first lens down to $d \sim$ 1.2 m$\phi$, which
we hope to fabricate using fused silica, without any significant 
loss of the image quality. 

\begin{table*}
\caption{List of selected mosaic CCD cameras.}
\label{mosaiccomparison}
\begin{center}
\begin{footnotesize}
\begin{tabular}{|l|cccc|ccc|c|c|c|}
\hline\hline
Camera & \multicolumn{4}{c|}{Telescope} & \multicolumn{3}{c|}{CCD} & FOV &
& First Light \\
Name & Name & $D$ [m] & $A$ [m$^{2}$] & $F$ & Vendor & Format & N$_{\rm CCD}$ &
$\Omega$ [deg$^{2}$] & $A\Omega$  & \\
\hline
WFPC2 & HST & 2.5 & 3.46& 12.9& Loral  &800$\times$800(15)& 3   & 0.0015&
0.005& Dec-93 \\ 
ACS & HST & 2.5 & 3.46& 12.9& SITe     &4k2k(15)& 2   & 0.0031&
0.011& Apr-02 \\ \hline
UH8K  & CFHT & 3.6 & 9.59& 4.2 & Loral  &4k2k(15) & 8   & 0.25  & 2.40& Sep-95 \\
SDSS  & SDSS & 2.5 & 3.83& 5  & SITe    &2k2k(24) & 30  & 6.0   & 22.99&May-98  \\
NOAO8K & Mayall & 3.8 & 9.98& 2.7 & SITe   &4k2k(15) & 8   & 0.36  & 3.59&
Jul-98$^*$ \\
CFH12K& CFHT & 3.6 & 9.59& 4.2 & MIT/LL &4k2k(15) & 12  & 0.375 & 3.60& Jan-99 \\
Suprime-Cam & Subaru & 8.2 & 51.65&2.0 & MIT/LL &4k2k(15) & 10  & 0.256 & 13.17&Jul-99
\\ \hline
MegaCam & CFHT & 3.6 & 9.59& 4.2 & Marconi&4.5k2k(13.5) & 40  & 1     & 9.59&
2002 \\
Pan-STARRS & Pan-STARRS & 2.0 & 2.14 & 4.0 & MIT/LL & 4k4k(12) & 64$\times$4$^\dag$ & 7$\times$4$^\dag$
& 15$\times$4$^\dag$ & 2006 \\
LSST$^\ddag$ & LSST & 8.4 & 37.40& 1.25& (TBD) &(1k1k(10))  & (1300)& (7.1)
& 265.54 & 2009 \\ \hline 
\multicolumn{11}{@{}l@{}}{\hbox to 0pt{\parbox{120mm}{\footnotesize
Notes. 
\par\noindent
\footnotemark[$*$] after the CCD upgrade to SITe'
\par\noindent
\footnotemark[$\dag$] Pan-STARRS is an array of telescopes. Initially, four telescopes will be built. 
\par\noindent
\footnotemark[$\ddag$] formerly called Dark Matter Telescope; based on a
report of a workshop, Science with the Large-aperture Synoptic
Survey Telescope, Tucson, Arizona, Nov 17-18, 2000. 
}\hss}}
\end{tabular}
\end{footnotesize}
\end{center}
\end{table*}

\section{Conclusion} \label{conclusion} 

We have designed and built a wide-field camera for the prime
focus of the 8.2-m Subaru telescope, which we named Suprime-Cam. 
Table~\ref{supsummary} shows the main characteristics of the camera.
Suprime-Cam is currently provided for open-use observations by Subaru
and has become one of the most frequently used instruments among others.
Nearly 30\% of all nights is allocated to the camera as of this
writing. This, we think, is because Suprime-Cam has an 
outstanding survey speed under sub-arcsec seeing, which is unique 
within the suite of instruments available on very large
telescopes. There has been no serious trouble which has stopped 
observing, although the burden of operations has not yet been quite 
minimized. 
For example, new filters require extensive trial-run for mechanical 
shakedowns. We will redesign some mechanics so as to reduce such 
operational costs. In the meantime, new cameras and telescopes 
are either being built or planned to offer much higher performance than 
we have achieved. We think that widening of the field of view 
is crucial for Suprime-Cam to remain competitive with such 
cameras in the future. 

\begin{table*}
\caption{Characteristics of Suprime-Cam.}
\label{supsummary}
\begin{center}
\begin{tabular}{|c|l|l|}
\hline\hline
            & Non-vignetted FOV & 30'$\phi$ (13.5 cm$\phi$) \\
Prime Focus & F-ratio & 1.86 \\
            & Field curvature & negligible \\
            & Atmospheric dispersion corrector & Lateral shift type
\\ \hline
            & Type & MIT/LL CCID20 \\
            & Format & 4096 $\times$ 2048 pixels \\
            & Pixel size & 15 $\mu$m (0.''202) \\
            & Number of CCDs & 10 (2$\times$5) \\
CCD         & Quantum efficiency & 400nm 55\% \\
            & (Typical)           & 600nm 80\% \\
            &                     & 800nm 89\% \\
            &                     & 1000nm 18\% \\
            & Operating temperature & -110$^{\circ}$C \\
            & Dark current        & 2-3 e hr$^{-1}$/pixel \\ 
            & FOV covered by mosaic & 34'$\times$27' \\
\hline 
            & Type            & Stirling cooler Daikin WE-5000  \\
Cooler      & Cooling capacity    & 5 W at 80 K,  16 W @163K \\
            & Overhaul maintenance cycle & 5000 h \\
\hline
Shutter     & Minimum exposure time & 1.2 s \\
\hline
Filter      & Physical dimension & 205 mm$\times$170 mm$\times$15 mm \\
            & Weight             & 1.3 kg \\
Filter exchanger & Number of loadable filters & 10 \\
            & Exchange time & 5 min \\
\hline
Size        & \multicolumn{2}{l|}{960 mm$\phi$$\times$ 1035 mm} \\
Weight      & \multicolumn{2}{l|}{295 kg}\\
Power consumption & \multicolumn{2}{l|}{420 W (600 W max)} \\ 
\hline
\end{tabular}
\end{center}
\end{table*}

\section*{Acknowledgments}

We are grateful to Prof. Katsuyuki Suzuki for his advice about the 
FEM simulation of the mechanical structure. We thank Wataru Kawasaki 
for his contribution in estimating the limiting magnitudes during 
the early phase of the project. We also appreciate Yasuhiro 
Sawada for his careful handcraft of wiring inside the dewar. 
Thanks are especially due to all of the staff members at 
Advanced Technology Center, National Astronomical Observatory 
for their help and support where Suprime-Cam was 
actually assembled and built. 

\appendix

\section{Effect of Atmospheric Differential Refraction} 

Atmospheric differential refraction affects various measurements
in astronomy. In particular, it often sets the ultimate limitation
on the field size we may hope to reach in wide-field imaging,
such as Schmidt observations (Bowen 1967; Wallace \& Tritton 1979;
Filippenko 1982; Watson 1984, and references therein).

The field of view of the Subaru prime focus is much smaller than
that of the typical Schmidt telescope. However,
atmospheric differential refraction may have some effects, 
even over a 30$'$ field of view of Suprime-Cam,
due to the very good image quality. We simulate the effect
based on the approximation formula derived by
Tanaka (1993), which is valid at the summit of Mauna Kea.

Tanaka (1993) expressed the refractive index, $n$, of the
atmosphere as $n=(1+\rho)$, and compute $\rho$ as follows.
At the sea level (pressure $p = 760$ mmHg), dry atmosphere with
0.03\% of CO$_2$ at the temperature $t=15^{\circ}$C have
\begin{equation}
\rho_2\times10^8=6432.8+\frac{2949810}{146-(1/\lambda)^2}
+\frac{25540}{41-(1/\lambda)^2},
\end{equation}
where $\lambda$ is the wavelength in units of $\mu$m
and the equation is valid for 0.2-1.35$\mu$ m.
This value is corrected for lower temperature $t$ ($^{\circ}C$)
and lower pressure $p$ (mmHg) at the summit as
\begin{equation}
\rho_1=\rho_2\times\frac{p_0[1+(1.049-0.0157t)\times10^{-6}p_0]}
{720.883\;(1+0.003661t)},
\end{equation}
where
\begin{equation}
p_0 = p\times\frac{g}{g_0},
\end{equation}
where $p$ is the measured pressure, $g$ is the measured gravity,
and $g_0=980.665$ gal is the standard gravity.
The effect of water vapor can be corrected as
\begin{equation}
\rho_0=\rho_1 - \frac{6.24-0.0680/\lambda^2}{1+0.003661t}
\times{f}\times10^{-8},
\end{equation}
with the pressure of water vapor $f$ (mmHg).
The atmospheric refraction can be computed by integrating the
deflection angle of incoming light,
\begin{equation}
di={\rm tan}\,i\;dn/n,
\end{equation}
from the observer, where $n=n_0=(1+\rho_0)$, to the
upper limit of the atmosphere.

Based on the model atmosphere for the set of
parameter values typical at the summit
of Mauna Kea (4180 m), that is, $p$ = 450mm Hg, $g$ = 978.627 gal,
$t$ = 0$^{\circ}C$, and $f$ = 1mmHg,
Tanaka (1993) obtained $\rho_0$ = 0.00017352 at $\lambda=$ 0.575 $\mu$m,
and derived the following approximation formula:
\begin{equation}
\label{eqn:app_eqn1}
r = 35.''746 \;{\rm tan}\,z - 0.''0431 \;{\rm tan}^3\,z
+ 0.''000164 \;{\rm tan}^5\,z,
\end{equation}
where $r$ is the atmospheric refraction and $z$ is the zenith
distance.
The differential refraction is also approximated as
\begin{equation}
\label{eqn:app_eqn2}
dr/dz = 0.''624\:{\rm sec}^2\,z \;\;({\rm deg}^{-1}).
\end{equation}

We assume that the guide star is at the field center, and
compute the positions of 8 hypothetical stars  projected
onto the standard coordinate ($\xi, \eta$) on the focal plane.
The offset from the optical axis (field center) of the eight
stars are $\Delta\alpha/{\rm cos}\,\delta=0', \pm15'$ and
$\Delta\delta=0', \pm15'$. The computation proceeds as follows.
Assume that the celestial sphere is a unit sphere, and let
the celestial North Pole and field center be $\vec{p}$ = (0, 0, 1)
and $\vec{s}_c=({\rm cos}\:\delta, 0, {\rm sin}\:\delta)$,
respectively.
Then, the zenith is expressed by
\begin{equation}
\vec{z}=({\rm cos}\,h\;{\rm cos}\,\phi, {\rm sin}\,h\;{\rm cos}\,\phi,
 {\rm sin}\,\phi),
\end{equation}
where $h$ is the hour angle, and $\phi$ is the latitude of the observer.
The zenith distance of the field
center is given by
\begin{equation}
{\rm cos}\:z_c = \vec{s}_c\cdot\vec{z}.
\end{equation}
This expression can be used to compute
\begin{equation}
{\rm tan}\:z_c=[1/{\rm cos}^2z_c -1 ]^{1/2},
\end{equation}
and then the atmospheric refraction $r_c$ according to equation
\ref{eqn:app_eqn1}.
U{\rm sin}g the thus-computed $r_c$, the coordinates of the refracted
field center are given by
\begin{equation}
\vec{s'}_c = \left( {\rm cos}\:r_c-\frac{{\rm sin}\:r_c}{{\rm tan}\:z_c}
 \right) \vec{s}_c + \frac{{\rm sin}\:r_c}{{\rm sin}\:z_c}\vec{z}.
\end{equation}
Similarly, a star located off the field center by ($\Delta\alpha,
\Delta\delta$) with
\begin{equation}
\vec{s}=[{\rm cos}\:\Delta\alpha{{\rm cos}(\delta+\Delta\delta)},
{\rm sin}\:\Delta\alpha{{\rm cos}(\delta+\Delta\delta)},
{\rm sin}(\delta+\Delta\delta)]
\end{equation}
is also refracted to
\begin{equation}
\vec{s'} = \left( {\rm cos}\:r-\frac{{\rm sin}\:r}{{\rm tan}\:z} \right)
\vec{s} + \frac{{\rm sin}\:r}{{\rm sin}\:z}\vec{z}.
\end{equation}

We then project $\vec{s'}$ onto $\vec{s'}_c$ using the
North Pole vector, $\vec{p}$, to obtain
\begin{equation}
\xi = \frac{1}{\vec{s'}\cdot\vec{s'}_c}
      \frac{(\vec{s'}_c\times\vec{p})}{|\vec{s'}_c\times\vec{p}|}
      \cdot\vec{s'},
\end{equation}
\begin{equation}
\eta = \frac{1}{\vec{s'}\cdot\vec{s'}_c}
      \frac{\vec{s'}_c\times(\vec{s'}_c\times\vec{p}))}
      {|\vec{s'}_c\times\vec{p}|} .
      \cdot\vec{s'}
\end{equation}

Figure~\ref{trail} shows the trail of 8 hypothetical stars
around the field centers of $\delta_c=-30^{\circ}$, $0^{\circ}$,
$+30^{\circ}$, and $+60^{\circ}$, respectively. The simulated exposure
was 6 hr from hour-angle $h=-3^h$ to $h=+3^h$
with a dot at every 0.5 hour.
Figure~\ref{maxexp} shows the maximum exposure time during which
the distortion of the stars at the edge of the
Suprime-Cam field does not exceed 0.''1.
Computations were made for 12 fields centered between
$\delta_c$= +80$^{\circ}$ and $\delta_c= -40^{\circ}$ with
$\Delta\delta_c=10^{\circ}$. For each field
13 exposures were simulated which start at hour angles
between $h=-3^h$ and $h=+3^h$ with $\Delta h =0.5$ hour.
The abscissa was the ${\rm sec}\,z$ at the beginning of
the exposure.

We make a few remarks based on this simulation. First,
if we took 0.''1 as the limit of distortion,
the longest allowable exposure time was about 40 min at
$z=30^{\circ}$ and about 15 min at $z=45^{\circ}$.
It is noted, however, that a longer exposure could be made depending
upon the location of the field. Second, even in the case we stacked
many short exposure frames, image distortion could become
a problem if individual exposures scatter in the wide range
of the zenith distances. Finally, actual guiding was provided using
a star near the edge of the field. The amount of the guiding
which gave the best image at the field center could be
computed by taking the effect of the differential refraction
into account.

\begin{figure}
\begin{center}
\FigureFile(80mm,80mm){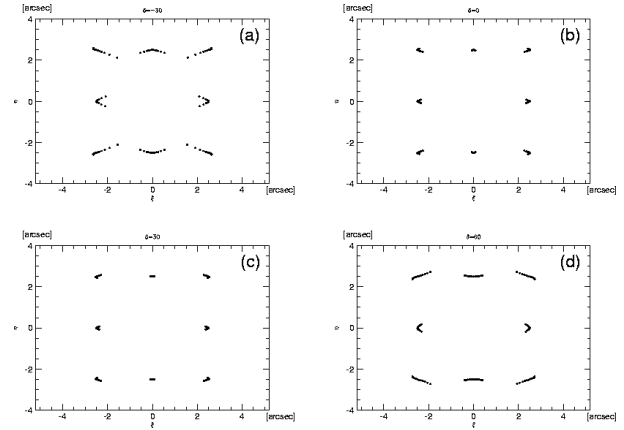}
\end{center}
\caption{
Trails of eight hypothetical stars showing the effect of the 
atmospheric differential refraction for the fields
centered at $\delta_c=$
$-30^{\circ}$ (a), $0^{\circ}$ (b),
$+30^{\circ}$ (c), and $60^{\circ}$ (d).
The stars are placed at $d\xi=0, \pm 15'$ , and
$d\eta=0, \pm 15'$ from the field center,
which correspond to the edge of the field of view of Suprime-Cam.
The simulated exposure was 6 hr from hour angle $h=-3^h$ to
$h=+3^h$ with a dot every 0.5 hour.
North is up and the position of each trail is arbitrarily shifted
so that the eight trails fit in one figure.}
\label{trail}
\end{figure}

\begin{figure}
\begin{center}
\FigureFile(80mm,80mm){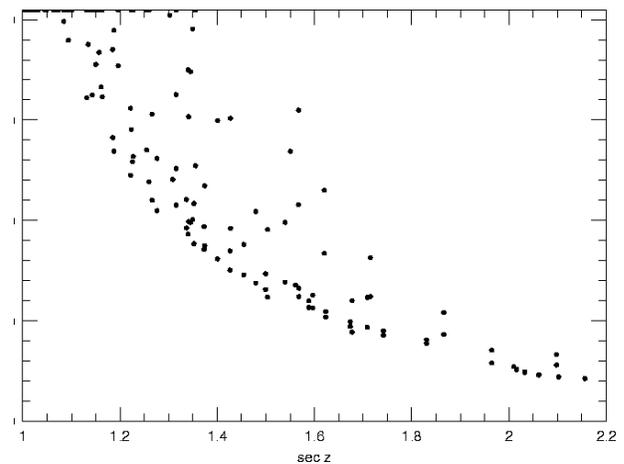}
\end{center}
\caption{
Maximum exposure time plotted as a function of the air mass
(${\rm sec}\,z$) at the beginning of an exposure.
The maximum exposure time is defined as the time
during which the distortion of stars at the edge of
the Suprime-Cam field does not exceed 0.''1.
Computations were made for 12 fields centered between
$\delta_c$= +80$^{\circ}$ and $\delta_c= -40^{\circ}$ with
$\Delta\delta_c=10^{\circ}$. For each field
13 exposures were simulated, which start at hour angles of
between $h=-3^h$ and $h=+3^h$ with $\Delta h =0.5$ hour.}
\label{maxexp}
\end{figure}

\end{document}